\documentclass[useAMS,usenatbib]{mn2e}
\usepackage{epsfig, color}

\newcommand{\ud}{\mathop{\mathrm{{}d}}\mathopen{}}

\title[Pop III multiplicity and dark star formation]
{Effect of Population III multiplicity on dark star formation}

\author[A. Stacy, A. H. Pawlik, V. Bromm and A. Loeb]
       {Athena Stacy$^{1,2}$\thanks{E-mail: athena.stacy@nasa.gov}, Andreas H. Pawlik$^{2}$, Volker Bromm$^{2,3}$ and Abraham Loeb$^{4}$\\
 $^{1}$NASA Goddard Space Flight Center, Greenbelt, MD 20771, USA \\
 $^{2}$Department of Astronomy and Texas Cosmology Center, University of Texas, Austin, TX 78712, USA \\
 $^{3}$Max-Planck-Institut f\"{u}r Astrophysik, Karl-Schwarzschild-Str. 1, 85741 Garching, Germany \\
 $^{4}$Astronomy Department, Harvard University, 60 Garden Street, Cambridge, MA 02138, USA}
\begin{document}

\pagerange{\pageref{firstpage}--\pageref{lastpage}}\pubyear{2011}


\maketitle
\topmargin-1cm

\label{firstpage}

\begin{abstract}
We numerically study the mutual interaction between dark matter (DM) and Population III (Pop~III) stellar systems in order to explore the possibility of Pop~III dark stars within this physical scenario.  We perform a cosmological simulation, initialized at $z \sim 100$, which follows the evolution of gas and DM.  We analyze the formation of the first minihalo at $z \sim 20$ and the subsequent collapse of the gas to densities of $10^{12}$ cm$^{-3}$.  We then use this simulation to initialize a set of smaller-scale `cut-out' simulations in which we further refine the DM to have spatial resolution similar to that of the gas.  We test multiple DM density profiles, and we employ the sink particle method to represent the accreting star-forming region.  We find that, for a range of DM configurations, the motion of the Pop III star-disk system serves to separate the positions of the protostars with respect to the DM density peak, such that there is insufficient DM to influence the formation and evolution of the protostars for more than $\sim 5000$ years.  In addition, the star-disk system causes gravitational scattering of the central DM to lower densities, further decreasing the influence of DM over time.  Any DM-powered phase of Pop III stars will thus be very short-lived for the typical multiple system, and DM will not serve to significantly prolong the life of Pop III stars.       
\end{abstract}


\section{Introduction}

The formation of the first stars was a crucial point in the early Universe.  Believed to have formed at $z > 20$ within $10^6$M$_{\odot}$ minihaloes (e.g. \citealt{haimanetal1996,tegmarketal1997,yahs2003}), these Population III (Pop III) stars began the process of transforming the Universe from a neutral and metal-free state to the ionized and metal-enriched state we observe today 
(e.g. \citealt{barkana&loeb2001,bromm&larson2004,ciardi&ferrara2005,glover2005,byhm2009,loeb2010}).  
However, the extent to which the first stars drove the evolution of the early Universe is highly dependent on their mass, luminosity, and effective temperature.  While the eventual mass of the first stars has usually been assumed to be driven by the cooling and chemistry of the baryons within early minihaloes, recent work has posed that dark matter (DM) may also affect the mass and evolution of Pop III stars beyond simply providing the initial gravitational potential well in which the star-forming gas first collapses.  For instance, \cite{spolyaretal2008,iocco2008} find that, if DM is composed of weakly interacting massive particles (WIMPs), energy from DM annihilation may also play an important role.

DM annihilation may first become significant during the initial collapse of gas within a minihalo.  This is due to the growing density of DM in the center of the minihalo as it responds to the growing potential well of the gas, generally termed `gravitational contraction.'  Several studies (e.g. \citealt{spolyaretal2008, freeseetal2008,natarajanetal2009}) find that gravitational contraction leads to sufficient DM annihilation to halt the collapse of the primordial cloud before a hydrostatic object has formed, leading instead to the formation of what has been termed a `dark star,' a giant ($\sim$ 1 AU) star powered by DM annihilation instead of nuclear burning.  

Because of the extended nature of these objects, their effective temperatures are too low to emit ionizing radiation.  Depending on how long gravitational accretion of DM continues, this may allow for a much longer gas accretion period before the dark star phase ends, after which the star begins contraction to the main sequence, and radiative feedback shuts off mass inflow onto the star (e.g. \citealt{spolyaretal2009}). \cite{ioccoetal2008}, 
find that the dark star phase is short-lived, 
though they did not follow the gradual accretion of gas onto the stars over time.  
More detailed work (\citealt{ripamontietal2010}) even finds that such `gravitational accretion' of DM during the initial cloud contraction will not halt the gas collapse, so a dark star of this type does not form.  Thus, the initial mass of the first stars will not be significantly affected.  However, other work reaches different conclusions (e.g. \citealt{freeseetal2008}).

At later stages, Pop III protostars may continue gathering DM through continued gravitational accretion as the protostar's mass and potential well grows.  They may also gain DM through `scattering accretion,' a process in which WIMPs scatter off the gas of the star and become gravitationally bound to it.  If the scattering cross section between WIMPs and baryons were large enough, and the resulting capture rate of DM by Pop III stars sufficiently high, this would prolong the lifetimes of Pop III stars.  This is because hydrogen will burn at a reduced rate while DM annihilation helps to support the star.  Recent work by \cite{sivertsson&gondolo2011}, however, find that this phase of scattering accretion will be very short-lived, $\la$ 10$^5$ yr, much smaller than the lifetime of the star.

However, the extent to which DM will alter the nature of Pop III stars depends highly on the DM density profile and the location of the DM density cusp with respect to the Pop III star 
Previous work has found that a single Pop III star will form at the center of a minihalo (e.g. \citealt{abeletal2002,brommetal2002,yoshidaetal2006}).  This is the picture that has been used by the previous studies of dark stars, and DM effects are likely to be strongest in this scenario.   However, recent work has found that Pop III stars do not necessarily form in isolation.  Instead of a single stellar peak at the center of a minihalo, a stellar multiple system is likely to form (e.g. \citealt{clarketal2008,turketal2009,stacyetal2010}; \nocite{clarketal2011a} Clark et al. 2011a, 2011b; \nocite{clarketal2011b} \citealt{greifetal2011,smithetal2011}; \nocite{stacyetal2011b} Stacy et al. 2011b).  In particular, the calculation by \cite{stacyetal2010}, which was started using cosmological initial conditions, found the formation of a Pop III multiple system within a rotating disk structure of order 1000 AU in radius.  
Recent calculations have furthermore found disk fragmentation and multiplicity to be robust against both photo-dissociating and ionizing feedback from the protostars (e.g. \citealt{smithetal2011}; Stacy et al. 2011b).  Even when allowing for stellar mergers on unresolved scales,  Stacy et al. (2011b) find that a massive binary remains at the end of their 5000 yr calculation.  This modified picture of Pop III star formation is also robust against statistical variation of the host minihaloes (\citealt{greifetal2011}).  

Only one-dimensional simulations have been used thus far to study the possible effects of DM annihilation on the first stars.  Three-dimensional effects, in particular stellar multiplicity and motion of the disk gas, may have important consequences for how DM and Pop III stars interact in the center of minihaloes
(see discussion in, e.g., \citealt{iocco2011}).  
One may initially guess that a DM profile could not remain peaked under the gravitational scattering effects of a star-disk system, and would thus be less prone to scattering accretion onto the stars.  As discussed in, e.g., \cite{sivertsson&gondolo2011}, however, the motion of a star through a DM halo may enhance its overall accretion rate, particularly for a lower-mass star that would need less DM to be powered as a dark star. 

To address these open issues, we perform a three-dimensional simulation which follows the evolution of a range of peaked DM density profiles in a minihalo that hosts a Pop III multiple system, yielding more accurate constraints on the influence of DM annihilation on the nature of the first stars, as well as the influence of a Pop III star-forming disk on the inner DM profile.   Representing each Pop III star that forms with a sink particle, we follow the gas and DM evolution for $2 \times 10^4$ yr ($\sim$ 500 free-fall times for our sink density $n=10^{12}$ cm$^{-3}$), and  we record the density of DM within the sinks as the evolution of the gas and DM profile is followed.   We initialize our simulation such that the maximum DM density is in the region of the most massive sink, giving an upper limit to the WIMP annihilation heating rate in this region as the star grows, as well as the potential for DM capture to prolong the life of the Pop III star.   

In Section 2 we describe the initialization of our simulations.  In Section 3 we present our results, including descriptions of the DM evolution, the Pop III growth rates, and the expected effects of DM on primordial gas and stellar evolution.  We conclude in Section 4.  

\section{Numerical Set-up}
The numerical simulations were run with {\sc gadget2}, a widely-tested three-dimensional N-body and SPH code (\citealt{springeletal2001,
springel2005}).    
We used adaptive gravitational softening for both gas and DM particles, but with an imposed minimum softening length (see Section 2.2).  The gravitational softening length was thus assigned to each particle based upon its current density, such that the softening length was adapted both throughout the simulation box and over time.   

A scale over which gravitational forces are `softened' is necessary when representing a smooth mass distribution with a finite number of simulation particles.  The ideal softening length $\epsilon$ to use in N-body and SPH simulations has been studied by a number of authors (e.g. \citealt{merritt1996,bate&burkert1997,athanassoulaetal2000}).  If $\epsilon$ is too small, the gravitational force between discrete N-body particles can become arbitrarily large, and computational timesteps prohibitively small.  Unphysically large fluctuations in force among the finite number of N-body particles will result.  If $\epsilon$ is too large, real features on scales smaller than $\epsilon$ will not be resolved (\citealt{merritt1996}).  
For SPH particles, collapse of Jeans-unstable clumps will be inhibited if $\epsilon$ is greater than the smoothing length.  On the other hand, setting $\epsilon$ to be less than the smoothing length may result in artificial fragmentation.  In a simulation with a large range of densities, the ideal $\epsilon$ will not be the same in both dense and diffuse regions.  Varying $\epsilon$ with the smoothing length of each particle provides a way to avoid the problems of $\epsilon$ values which are too large or small (e.g. \citealt{price&monaghan2007,iannuzzi&dolag2011}).  This does lead to some extra computational cost in our simulations, however, as full adaptive softening requires smoothing kernels to be calculated not only for the SPH particles but also the DM.

Our simulations evolve the central regions of the first minihalo that formed in a previous, larger-scale cosmological simulation.   We furthermore replace the DM from the minihalo with more highly refined particles, assuming a range of density profiles.  The gas from this region is already resolved down to scales of 50 AU.  As found in, e.g., \cite{spolyaretal2008, natarajanetal2009,ripamontietal2010}, our resolution length is the approximate scale at which the DM annihilation heating rate may first surpass the gas cooling rate and prevent gas collapse to stellar densities, given a sufficiently peaked DM profile that is aligned with a single stationary star.  Our simulations thus have sufficient gas and DM resolution to determine whether this may still be the case within a Pop III star-disk system.
In the following sections we describe in further detail our numerical initialization.    

\begin{figure*}
 \includegraphics[width=.45\textwidth]{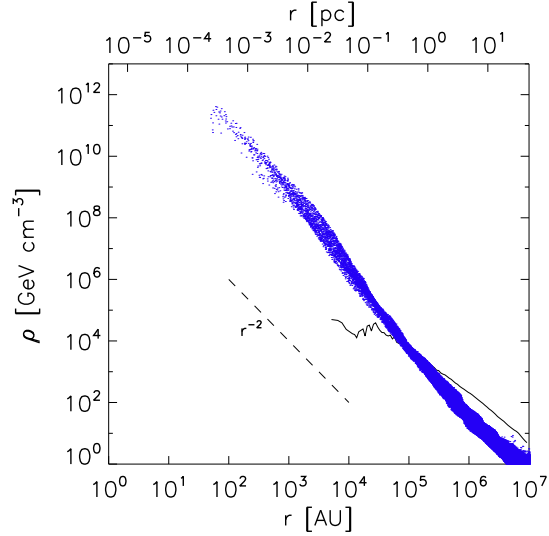}
\caption{Density of gas and DM as a function of distance from most massive sink, as measured from the initial cosmological simulation.  Recall that 1 GeV  cm$^{-3}$ corresponds to $1.7 \times 10^{-24}$ g cm$^{-3}$.  Blue dots represent individual gas particles, which roughly follow an $r^{-2}$ density profile (dashed line).  Solid line denotes the radially averaged DM profile to which the more refined small-scale simulations were normalized.  }
\label{dens_orig}
\end{figure*}

\begin{figure*}
 \includegraphics[width=.45\textwidth]{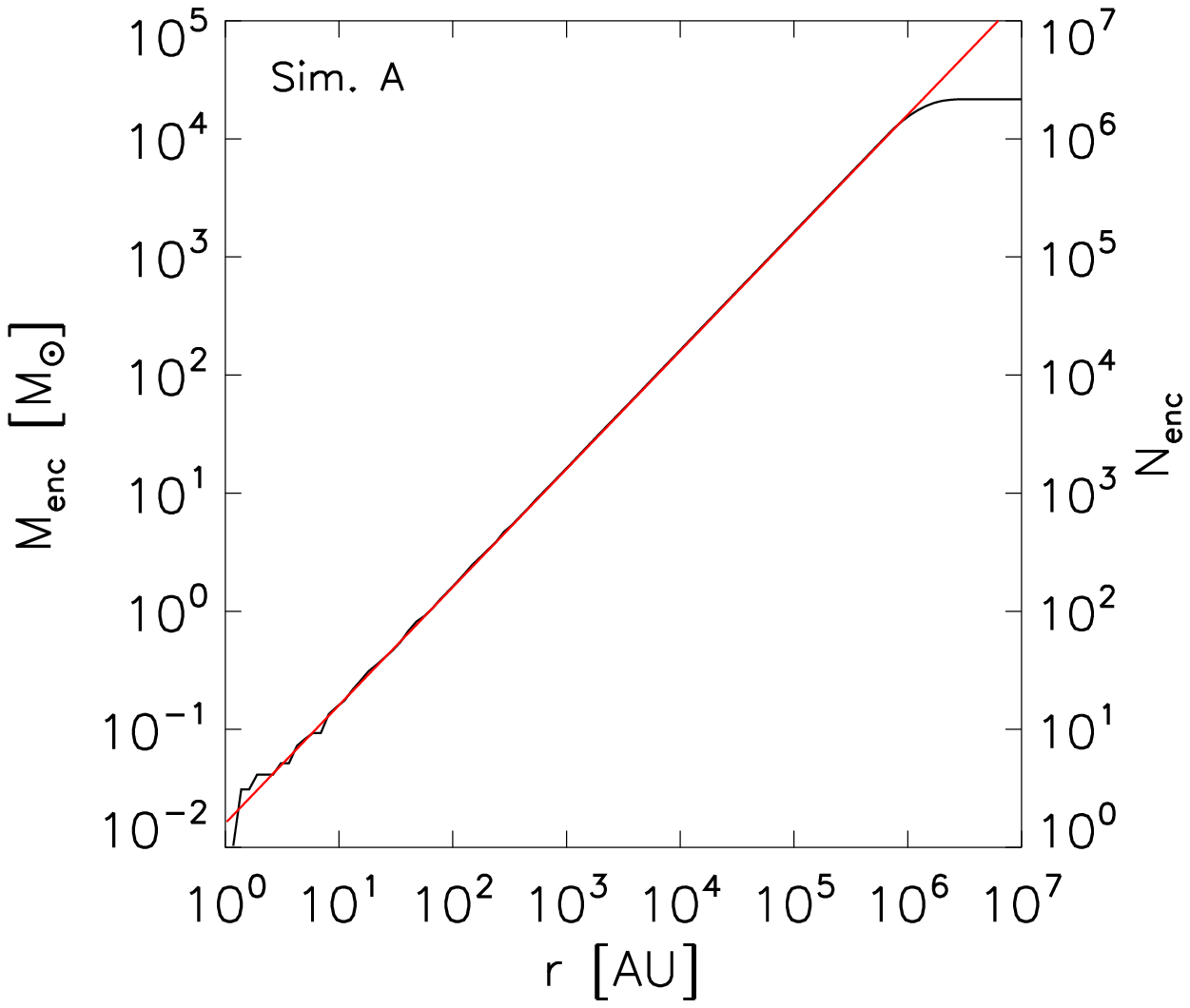}
 \includegraphics[width=.45\textwidth]{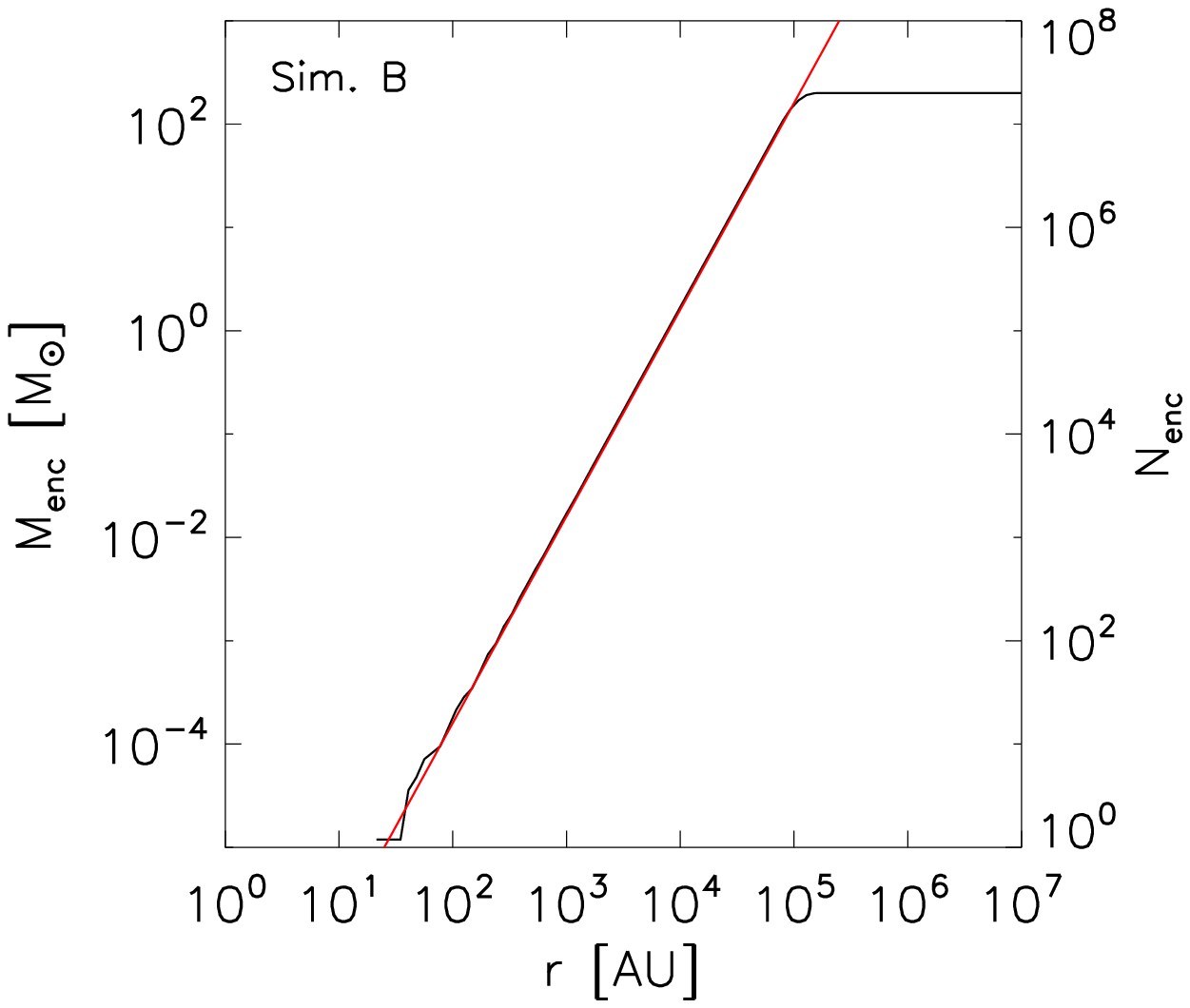}
\caption{ {\it Left:} Enclosed mass $M_{\rm enc}$ versus radius after refinement for Simulation A ($\rho_{\rm DM} \propto r^{-2}$).  {\it Right:}  Enclosed mass versus radius for Simulation B ($\rho_{\rm DM} \propto r^{-1}$).   Red lines show the $M_{\rm enc}$ profile to which the DM was initialized.}
\label{menc_init}
\end{figure*}

\subsection{Gas Initialization}
 The gas distribution and its thermal and chemical state were taken from a snapshot from a previous simulation of Pop III star formation (\citealt{stacyetal2010}).  This earlier simulation was initialized at $z=99$ in a 100 $h^{-1}$ kpc (comoving) periodic box.   This was done in accordance with a $\Lambda$CDM cosmology
with $\Omega_{\Lambda}=0.7$, $\Omega_{\rm M}=0.3$, $\Omega_{\rm B}=0.04$, and $h=0.7$.  To accelerate structure formation, we used an artificially enhanced normalization of the power spectrum, $\sigma_8=1.4$.  

This previous simulation employed a standard hierarchical zoom-in procedure to increase the resolution of the central 20 comoving kpc (see \citealt{stacyetal2010} for further details).  
The most highly resolved DM particles had masses of  $m_{\rm DM} =0.096$~M$_{\odot}$.  Thus, scales of $\la$ 10$^4$ AU which enclosed $\la$ 1 M$_{\odot}$ of DM ($\sim$ ten DM particles) were not well-resolved (see Fig. \ref{dens_orig}).  
The most highly resolved gas particles had a mass $m_{\rm SPH} =0.015$~M$_{\odot}$, giving a mass resolution of $M_{\rm res}\simeq 1.5 N_{\rm neigh} m_{\rm SPH} \la  1$ M$_{\odot}$, where $N_{\rm neigh}\simeq 32$ is the typical number of particles in the SPH smoothing kernel (e.g. \citealt{bate&burkert1997}).  The gas evolution was then followed until it reached a maximum density of 10$^{12}$ cm$^{-3}$, at which point we implemented the `sink particle' method.  This is a computational technique in which a small, high-density, gravitationally collapsing region is replaced with a single sink particle that will grow in mass as gas particles continue to fall onto this region.  This circumvents the need to continue following the evolution of the star-forming region to yet higher densities, which is very computationally expensive, and allows for a numerical representation of a growing Pop III star. Our particular implementation of the sink particle method is identical to that in \cite{stacyetal2011}.  In short, we check each gas particle to determine if it satisfies the criteria  $d < r_{\rm acc}$ and $j_{\rm SPH} < j_{\rm cent}$, where  $d$ is  the distance of the particle relative to the sink, $ r_{\rm acc}\simeq 50$ AU is the accretion radius, $j_{\rm SPH} = v_{\rm rot} d$ is the angular momentum of the gas particle,  $v_{\rm rot}$ is the gas particle's rotational velocity, $j_{\rm cent} = \sqrt{G M_{\rm sink} r_{\rm acc}}$ is the angular momentum required for centrifugal support, and $M_{\rm sink}$ is the mass of the sink particle.
If these criteria are satisfied, the gas particle is removed from the simulation, and the mass of the accreted particle is added to that of the sink.  Our sink accretion algorithm also allows for the merging of two sink particles if the smaller sink satisfies the criteria listed above.  

The gas of the central star forming region is taken from this previous simulation at the point immediately after the first sink has formed.  The gas density profile at this point can be seen in Figure \ref{dens_orig}.  Depending on the particular DM density profile, we cut out a central cube that is either 1 or 20 pc on a side, equivalent to a total of 3.6$\times 10^4$ gas particles (540~M$_{\odot}$) and 2.5$\times 10^5$ gas particles (4000~M$_{\odot}$), respectively.  Note our cut-out regions contain only the most highly-resolved gas particles from the original cosmological simulation, so every gas particle has a mass of $0.015$~M$_{\odot}$.  
We then resimulate only the cut-out region, excluding the outer less-resolved regions that were in the full-scale cosmological simulation.   The new cut-out simulations have the same mass resolution for the gas ($\la  1$ M$_{\odot}$) and use the same sink particle technique,  but now include the addition of a peaked DM profile that is resolved on sink scales and centered on the most massive sink.

Note that the gas in the outer parts of the cosmological simulation can be safely ignored for our purposes due to the long dynamical times.  The gas at the edge of the cut-out has typical densities of 10$^2$-10$^4$ cm$^{-3}$, corresponding to free-fall times of $\sim 5 \times10^5$ to $\sim 5 \times10^7$ yr.  This is over ten times longer than the length of time over which we evolve the gas in the cut-outs (20,000 yr).  Furthermore, a rarefaction wave at the cut-out edge could possibly develop due to the vacuum boundary conditions.  However, such a wave will only travel a distance of  $c_{\rm s} \, t \sim 8500$ AU, where $c_{\rm s}$ is the gas soundspeed ($\sim$ 2 km s$^{-1}$), and the time $t$ is 20,000 yr.  This is less than one-tenth of the half-length of our smallest cut-out.

By adding the highly resolved DM only after the first sink has formed, we are assuming that DM annihilation did not have a significant effect on the gas evolution at low densities and was unable to halt the gas collapse before it reached densities of 10$^{12}$ cm$^{-3}$.  This is consistent with the results of \cite{ripamontietal2010}, who find that up to $n \sim 10^{14}$ cm$^{-3}$ the evolution of temperature with density will not vary significantly for a range of DM parameters.   They furthermore find that heating through DM annihilation surpasses cooling at critical densities of $n_{\rm crit} \sim 10^9 - 10^{13}$ cm$^{-3}$, similar to the critical densities found by \cite{spolyaretal2008, natarajanetal2009}.  However, \cite{ripamontietal2010} find that this will not halt the continued collapse of the gas, with the exception of a very brief ($\sim$ 3yr) stall in collapse for their most extreme combination of DM parameters.  Instead, the excess heat goes mainly into dissociating H$_2$, and the continuum-dominated cooling regime begins earlier.  Nevertheless, their work was a one-dimensional model, and future work should confirm that their results would hold in a three-dimensional study.     
For our current study, however, we do not directly follow the gas and protostellar evolution on these sub-sink scales, and instead aim to constrain the total DM reservoir available within the sink region of 50 AU over the initial accretion phases of the Pop III protostars.
With these uncertainties in mind, our set-up also  allows us to find an upper limit to the mass reached by the Pop III stars under various DM conditions, since we do not include the feedback effects of protostellar radiation on the surrounding gas (e.g. Smith et al. 2011; Stacy et al. 2011b).   

\subsection{DM Initialization}

\subsubsection{Density Initialization}

The DM, which was unresolved on sink particle or stellar disk scales in \cite{stacyetal2010}, was initialized with two different density powerlaws , $\rho_{\rm DM} \propto r^{-2}$ (`Simulation A') and  $\rho_{\rm DM} \propto r^{-1}$ (`Simulation B'), with the normalization determined by the total mass of DM found in the corresponding region of the original simulation.  The corresponding profiles of enclosed mass $M_{\rm enc}$ can be seen in Figure \ref{menc_init}.

The density profiles were generated starting from an initial uniform
density field. This field was generated by placing particles at
glass-like positions, which was achieved by allowing randomly placed
particles to evolve under an artificial negative gravitational force
until a quasi-equilibrium configuration is reached
(\citealt{white1996}).  Note that this method avoids small-scale fluctuations in the relative distances between particles.  This is an improvement upon a Monte Carlo sampling of the density field, which would be subject to such Poisson noise.

The particles of uniform density $\rho_0$ can then be transformed to a powerlaw density 
$\hat{\rho}  \propto \hat{r}^{-n}$ through the coordinate transformation $(r,\theta,\phi) \rightarrow (\hat{r},\theta,\phi)$. The new coordinates will satisfy

\begin{equation}
\hat{\rho}(\hat{r}) \hat{r}^2 {\rm sin}\theta \ud \hat{r}   \ud \theta \ud \phi =   \rho_0 r^2 {\rm sin}\theta \ud r \ud \theta \ud \phi \mbox{,}
\end{equation}

\noindent from which we can derive the relation

\begin{equation}
\hat{r} \propto r^{3/(3-n)} \mbox{, for \ }
0 \le n < 3 \mbox{.}
\end{equation}

We initially align the DM density peak with the gas density peak, represented by the first sink particle.  Figure \ref{dens_orig} shows the initial DM density profile, taken from the original full-scale simulation, used to normalize the DM mass in our current calculations.  
We use density units of 1 GeV  cm$^{-3}$, which corresponds to $1.7 \times 10^{-24}$ g cm$^{-3}$.

We note that our DM initialization also accounts for gravitational contraction on scales greater than $\sim$ 10$^4$ AU (0.05 pc) during earlier phases of the central gas and DM evolution.  This is
because we normalize the DM mass based on the central DM content built up in the full-scale simulation only after the gas has collapsed to its 
maximum resolvable density.  Further gravitational contraction on scales less than $10^4$ AU is not resolved, however, so Simulations A and B are initialized to cover a range of possible DM configurations on small scales
in order to follow the continued gravitational accretion of DM onto the protostellar regions.  

For Simulation A, we use $128^3$ DM particles, and the mass of each DM particle is slightly smaller than the gas,  $m_{\rm DM} =0.01$~M$_{\odot}$.  We thus normalized the DM density profile to provide a total of $\sim 2 \times 10^4$ M$_{\odot}$ of DM enclosed within a 20 pc cube, the same DM mass found in the corresponding region of the original cosmological simulation.  On these scales, the DM thus dominates over the 4000 M$_{\odot}$ of gas enclosed within the same region.  We also impose a minimum softening length of 10 AU.  
For Simulation B, we used $256^3$ DM particles, each having a mass of $m_{\rm DM} = 1 \times 10^{-5}$~M$_{\odot}$.   
This yielded a total DM mass of $\sim 200$ M$_{\odot}$ inside a 1 pc cubical region, the same DM mass that resided in the central 1 pc of the original cosmological simulation.  
On scales of 1 pc it is the total mass of gas, 540 M$_{\odot}$, that dominates over the DM.   However, were the Simulation B profile extended to encompass a larger 10 pc cube, the total enclosed mass would be a factor of 100 greater, $\sim 2 \times 10^4$ M$_{\odot}$, and DM would dominate the gas on larger scales.   In comparison, the same volume in Simulation A holds a similar mass of DM, $\sim 1 \times 10^4$ M$_{\odot}$.  For Simulation B we use a minimum softening length of 25 AU.  

A larger particle number was required for this shallower profile so that the DM could be resolved on scales less than 50 AU while still extending out to distances greater than the stellar disk, which has a radius on the order of 1000 AU.  Note that for Simulation B, where $\rho_{\rm DM} \propto r^{-1}$, the enclosed mass increases as $M_{\rm enc} = m_{\rm DM}N_{\rm part} \propto r^{2}$, where $N_{\rm part}$ is the total number of DM particles used in the simulation.  We place approximately 15 particles within the central 100 AU of Simulation B (Fig. \ref{menc_init}).  Since $N_{\rm part} \propto r^{2}$, this means that a box size that is 1000 times larger, $10^5$ AU, will require a factor of $10^6$ more particles, $N_{\rm part} \sim 1.5\times10^7$.  This is already close to the $256^3$ particles available for Simulation B, so the half-length of the box in this case does not extend beyond  $10^5$ AU (= 0.5 pc), or a total length of $2 \times 10^5$ AU (= 1 pc).  For this reason Simulation B uses a box size of 1 pc.

For the $\rho_{\rm DM} \propto r^{-2}$ profile of Simulation A, however,  we have $N_{\rm part} \propto r$.  In this case, the inner region is better-resolved because we now have $\sim$ 30 particles within only 20 AU.  To resolve scales $10^5$ times larger than this now requires a factor of $10^5$ more particles, or $3 \times 10^6$ particles within $2 \times 10^6$ AU (= 10 pc).  A box with this half-length will then use all of the available $128^3$ particles, and for Simulation A the full length of the box is thus $4 \times 10^6$ AU (= 20 pc).    

We test multiple DM profiles since, on these very small scales ($\la$ 1 pc), the DM density is still a matter of uncertainty.  Many numerical studies have predicted that an inner $r^{-1}$ DM cusp will form within galactic haloes (e.g. \citealt{navarroetal1996}), and that the total star+DM mass within the central regions will remain stable to subsequent star formation and mergers (e.g., \citealt{loeb&peebles2003,gaoetal2004}). However, analytic and numerical models of DM evolution during the cooling and infall of baryons find that adiabatic contraction will further increase the central DM densities (e.g., \citealt{blumenthaletal1986,gnedinetal2004}).  This is in contrast to observations of galaxies that point to central DM concentrations that are flat instead of peaked (the widely-known `cusp-vs.-core' problem, e.g. \citealt{moore1994,burkert1995,debloketal2001,gentileetal2005,spekkensetal2005,battagliaetal2008}).  The innermost resolved DM of our initial cosmological simulation in fact shows a profile of approximately $\rho_{\rm DM} \propto r^{-1.5}$ (see Fig. \ref{dens_orig}).  Our simulations therefore cover a range of these possibilities.

\subsubsection{Velocity Initialization}
 
For a given spherically symmetric density profile, an isotropic distribution function (DF) for the DM particles can be generated using Eddington's formula (\citealt{binney&tremaine2008}, Equation 4.46):

\begin{equation}
f(\mathcal E)=\frac{1}{\sqrt{8}\pi^2}\left[\int^{\mathcal E}_0 \frac{\ud \Psi'}{\sqrt{\mathcal E-\Psi'}}\frac{\ud^2\rho}{\ud\Psi'^2}+\frac{1}{\sqrt{\mathcal E}}\left(\frac{\ud\rho}{\ud \Psi'}\right)_{\Psi'=0}\right] 
\end{equation}
 
\noindent where $f(\mathcal E)$ is the distribution function in units of mass per phase space volume, e.g., g cm$^{-3}$ (cm s$^{-1}$)$^{-3}$, and $\Psi$ is the relative potential, which can be set to the negative of the gravitational potential as measured from the edge of the system.  $\rho$ is the mass density of the system at the given $\Psi$, and $\mathcal E$ is the relative energy of DM per unit mass.
If $\Psi$ and the relative energy $\mathcal E$ are known for a particle, then its velocity $v$ can be calculated using


\begin{equation}
v = \sqrt{2(\Psi - \mathcal E)}
\end{equation}

For the input density we assumed the DM profile is cut off at an outer radius of 10 pc for Simulation A and 0.5 pc for Simulation B.  We further assume the profile flattens to a uniform-density core in the central 5 AU, well inside the extent of the sink particles. 
To assign velocities to each particle, we first divided the DM profile into 3000 radial bins, centered upon the densest particle.  We then calculated the relative potential $\Psi$ at each bin.  Each particle was assigned the value for $\Psi$ corresponding to its bin, and a value for its relative energy $\mathcal E$ was randomly drawn from the DF.  The amplitude of the velocity $v$ was then found using Equ. 4.
In the Appendix we provide more details about randomly drawing a value of $\mathcal E$ from the distribution function $f(\mathcal E)$ .

\begin{figure*}
 \includegraphics[width=.4\textwidth]{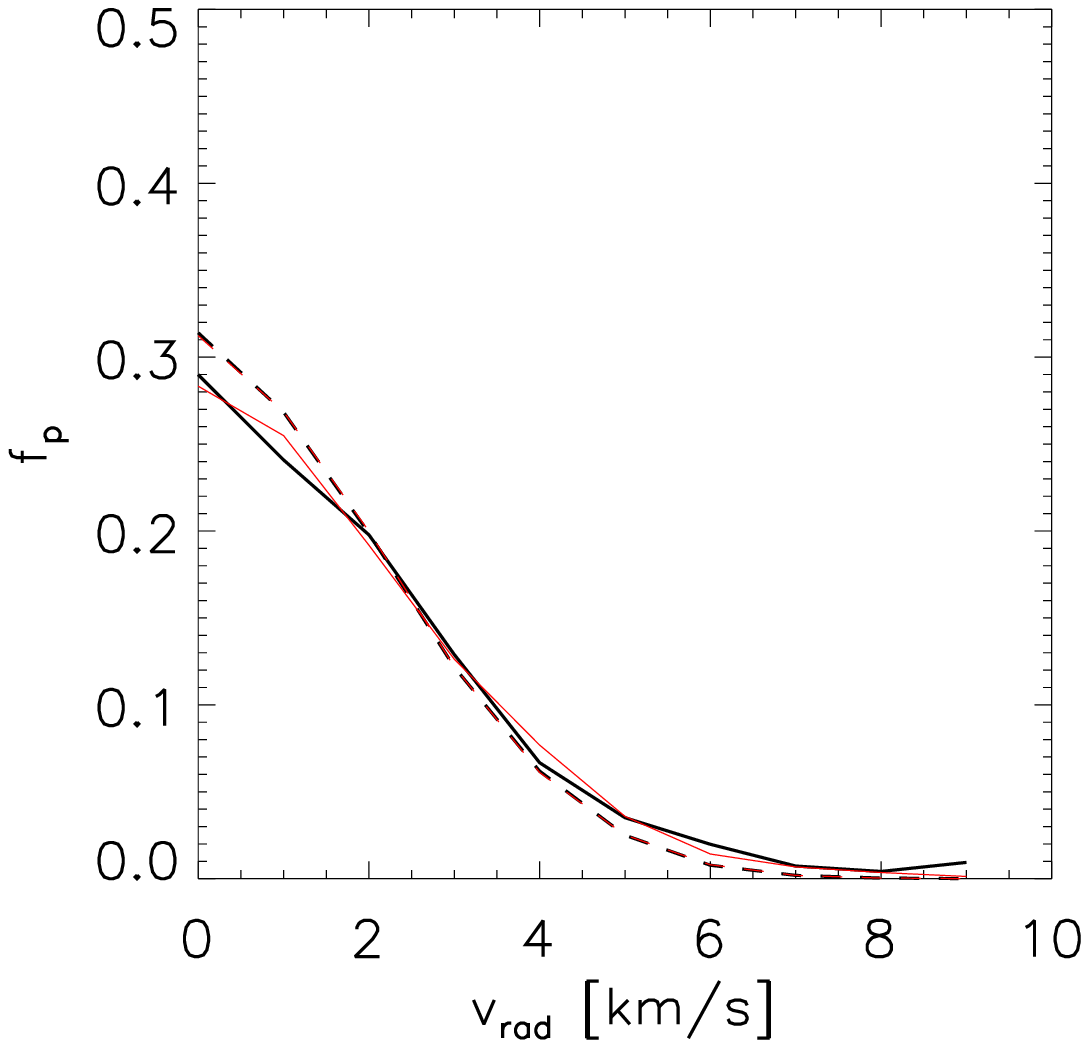}
 \includegraphics[width=.4\textwidth]{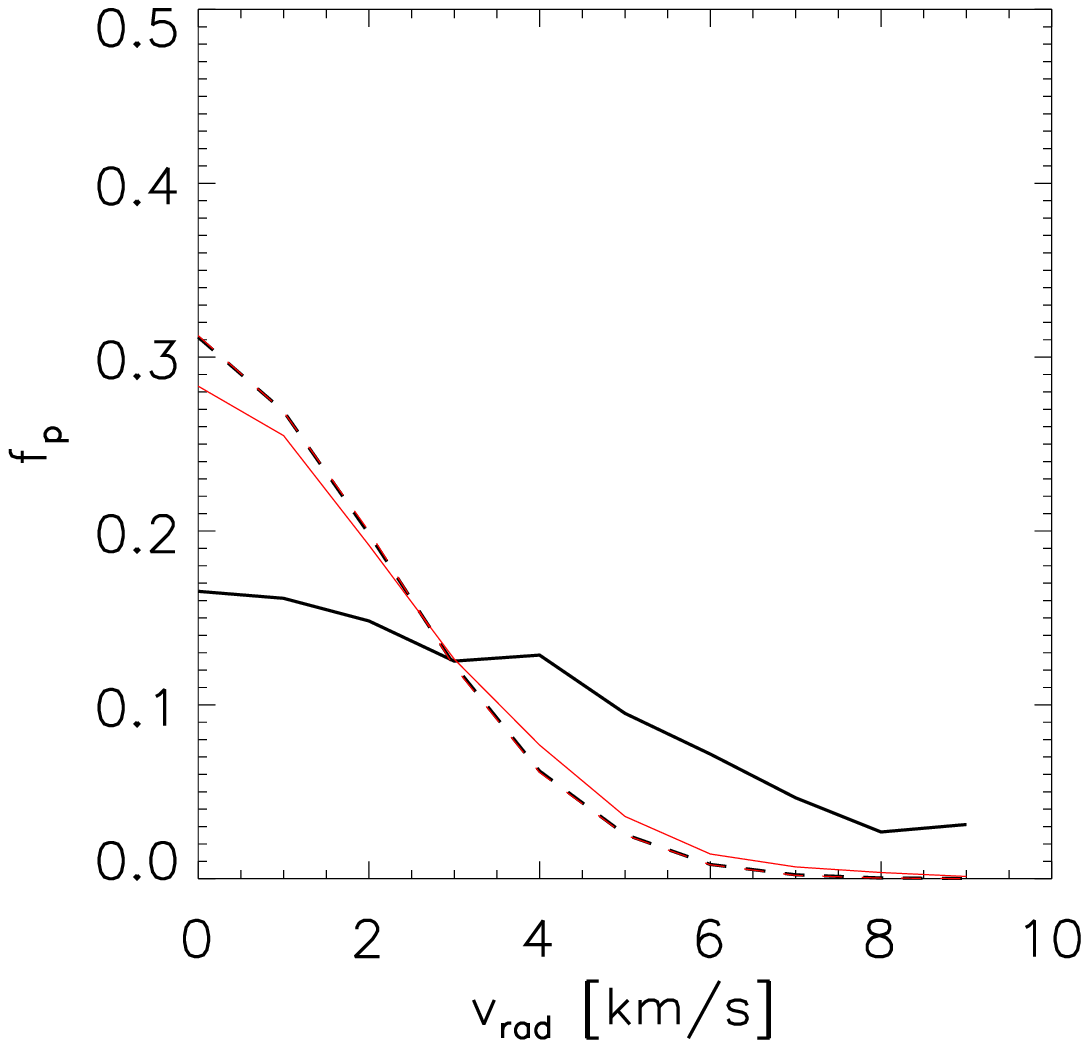}
 \caption{Velocity distribution of DM at the end of Simulation A. Solid line is for particles inside 10$^4$ AU from the central DM peak.  Dashed line is for particles greater than 10$^4$ AU from the peak.  For comparison, the corresponding red lines show the velocity distribution to which the DM was initialized.  {\it Left}: `No-gas' case.   {\it Right:} `With-gas' case.  While the velocity distribution is stable for greater than 5$\times$10$^4$ yr ($\sim$ 1000 free-fall times for typical peak DM density $\rho_{\rm DM}=10^{12}$ GeV cm$^{-3}$) in the `no-gas' case, the interaction between the central DM and the stellar disk in the `with-gas' case has greatly flattened the velocity distribution within 10$^4$ AU, while little change occurs beyond the disk at distances greater than 10$^4$ AU. }
\label{vel_rm2}
\end{figure*}

\begin{figure*}
 \includegraphics[width=.4\textwidth]{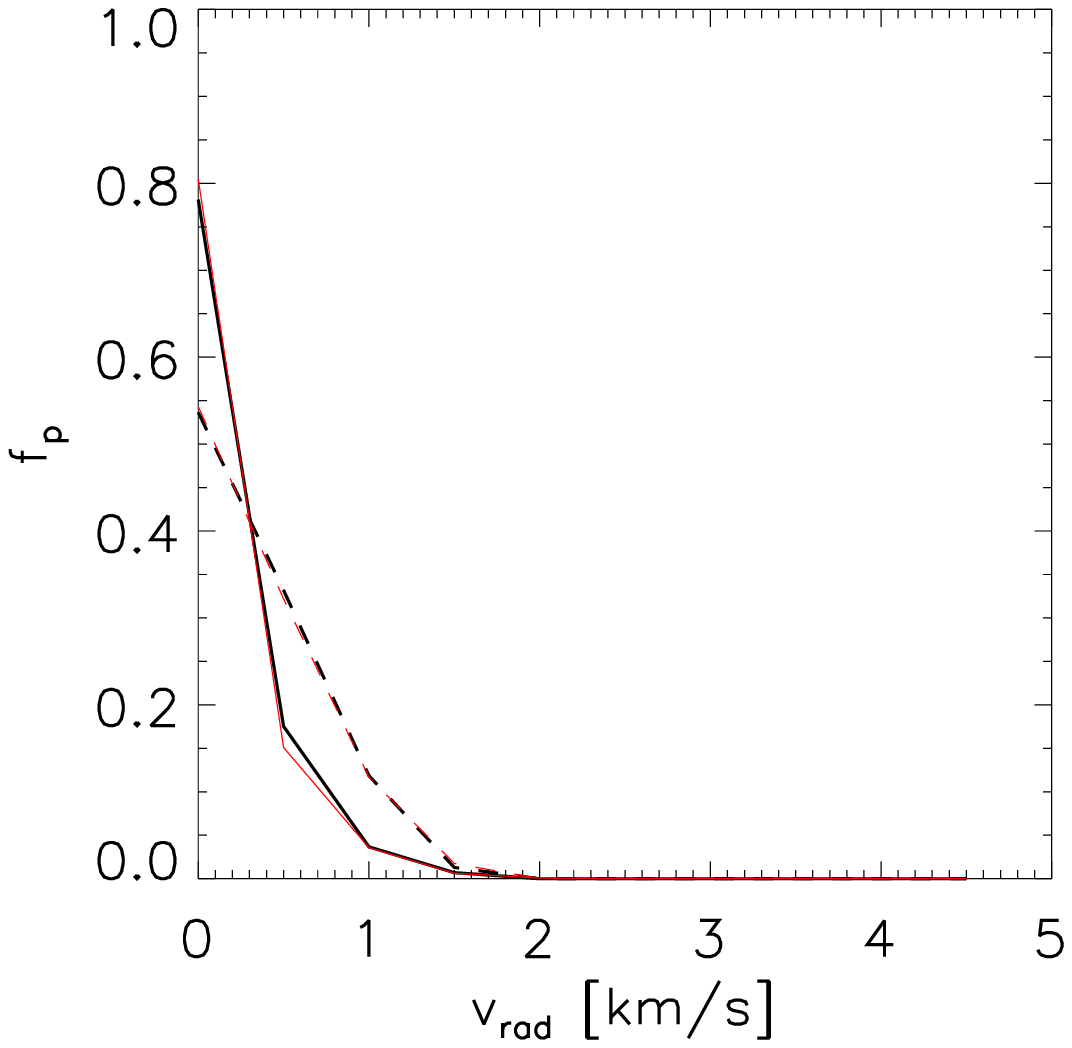}
 \includegraphics[width=.4\textwidth]{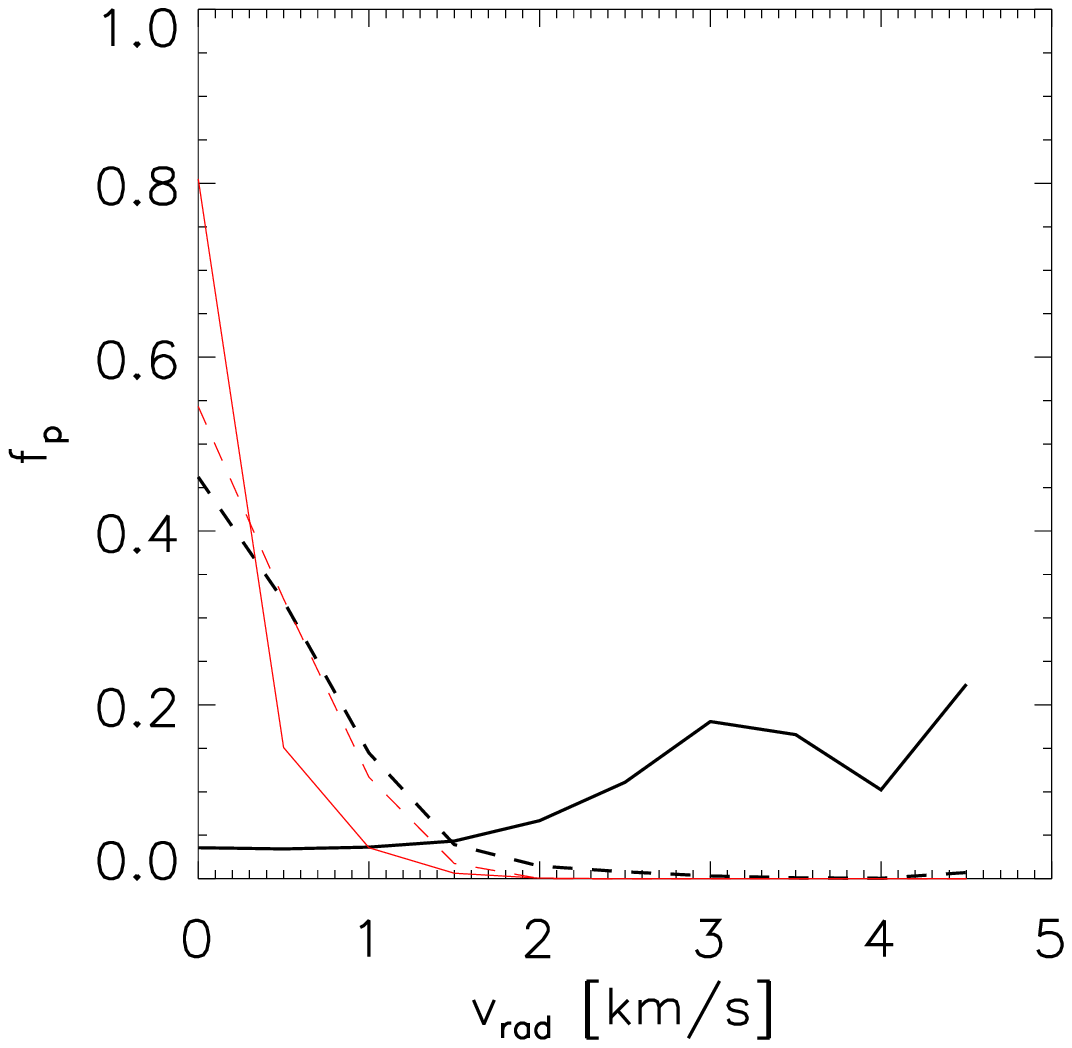}
 \caption{Velocity distribution of DM at the end of Simulation B. Notation is the same as in previous figure.  {\it Left}: `No-gas' case.   
 {\it Right:} `With-gas' case.  The velocity distribution is stable for greater than 5$\times$10$^5$ yr ($\sim$ 100 free-fall times for typical peak DM density $\rho_{\rm DM}=10^{8}$ GeV cm$^{-3}$)  in `no-gas' case.  Similar to Simulation A, the interaction between the DM and the stellar disk in the `with-gas' case has greatly flattened the inner velocity distribution. 
}
\label{vel_rm1}
\end{figure*}

To determine the velocity component along each Cartesian axis ($v_x$, $v_y$, and $v_z$), we next picked two angles at random, $\theta$ and $\phi$ (e.g. \citealt{widrow2000}).  Given two random numbers $p$ and $q$ ranging between 0 and 1, we set  $\theta = {\rm {cos^{-1}} (1 - 2{\it p})}$ and  $\phi = 2\pi q$  We used these angles to map each particle in velocity space, where $v_x = v \rm \, sin\theta\, cos\phi$, $v_y = v \rm \, sin\theta\, sin\phi$, and $v_z = v \rm \, cos\theta$.  The resulting radial velocity distribution can be seen in Figures \ref{vel_rm2} and \ref{vel_rm1}.

\section{Results}

\subsection{DM Evolution}

 We describe results for two sets of simulations: `Simulation A', in which the DM was initialized with $\rho_{\rm DM} \propto r^{-2}$, and `Simulation B', which was initialized with $\rho_{\rm DM} \propto r^{-1}$.   
 
 We first describe Simulation A.  We initially evolve Simulation A with only DM for $\sim$ 50,000 yr ($\sim$ 1000 free-fall times for typical peak DM density $\rho_{\rm DM}=10^{12}$ GeV cm$^{-3}$) .  The resulting DM configuration is then used as the initial conditions for the `with-gas' case, in which the gas and initial sink are placed in the center of the DM minihalo and evolved for a total sink accretion time of $t_{\rm acc} \sim 20,000$ yr.  In the `with-gas' case, several more sinks form and a disk $\sim$ 1000 AU develops.  For comparison, we also evolve the same halo for another 20,000 yr without gas (the `no-gas' case). 

\begin{figure*} 
\includegraphics[width=.4\textwidth]{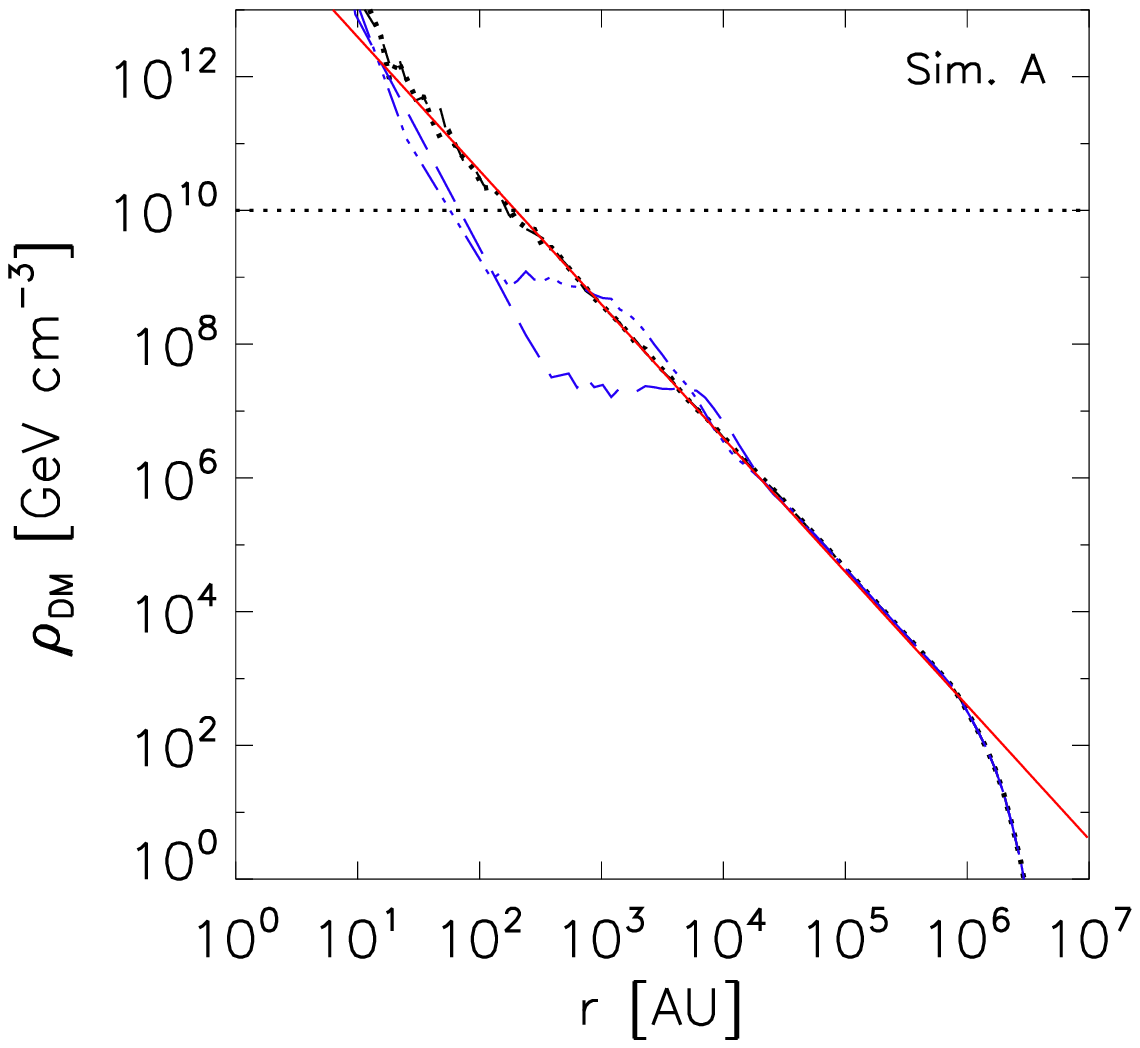}
\includegraphics[width=.4\textwidth]{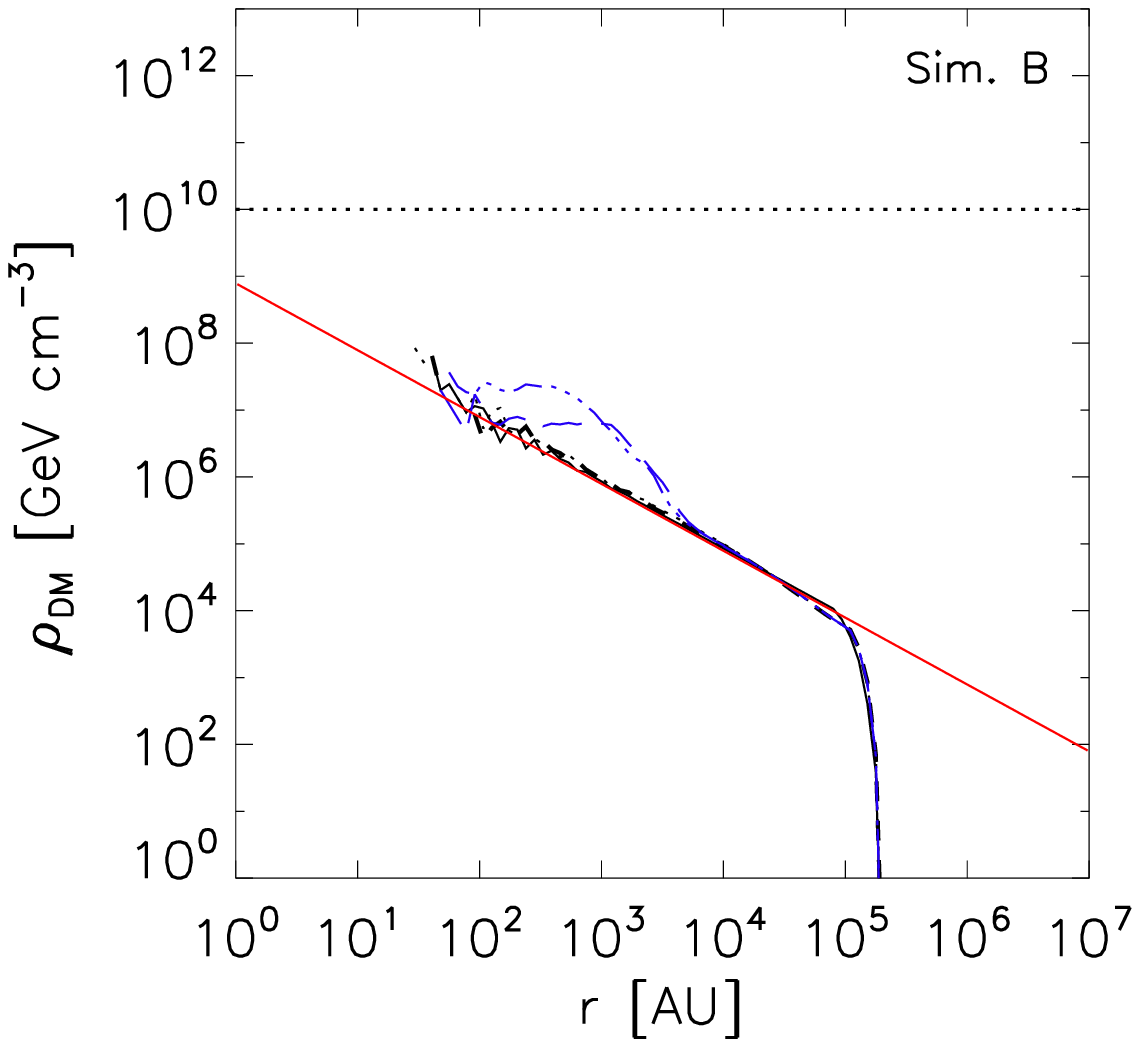}
\caption{Evolution of density structure of DM 
as measured from the densest DM particle. 
Red lines are analytic fits the to density profile to which the DM was initialized.  Black lines represent `no-gas' cases, while blue lines represent `with-gas' cases.  
Dotted lines represent typical density threshold for DM scattering accretion to support a star, given $m_{\rm wimp} = 100$ GeV
(Freese et al. 2008b, Yoon et al. 2008, Iocco 2008, Iocco et al. 2008, Spolyar et al 2009).
 {\it Left}: Simulation A.  The `no-gas' case is shown at 50,000 yr (black dotted line), and after a further 20,000 yr (black dashed line).   Also shown is the DM density for the `with-gas' case after $t_{\rm acc} =$ 10,000 yr (blue dashed triple-dot line) and  $t_{\rm acc} =$ 20,000 yr (blue long-dashed line). {\it Right:} Simulation B at 5 $\times$ 10$^5$ yr (black dotted line), and after a further 20,000 yr (black dashed line).  `With-gas' is shown at $t_{\rm acc} =$ 10,000 yr (blue dashed triple-dot line) and  $t_{\rm acc} =$ 20,000 yr (blue long-dashed line).  The central regions are stable in each `no-gas' case.  However, the corresponding regions in the `with-gas' case of Simulation A has a lower mass and density due to scattering by the star-disk system over 20,000 yr.  In contrast, the `with-gas' case of Simulation B sees a temporary increase in central DM density, but this increase is insufficient to affect the evolution of the primordial gas, and by 20,000 yr $\rho_{\rm DM}$ has decreased again.  }
\label{dens_rm2}
\end{figure*}

Figure \ref{vel_rm2} shows how the distribution of the magnitude of DM radial velocity evolves for the `no-gas' case compared to the `with-gas' case for Simulation A, while the corresponding density evolution can be seen in Fig. \ref{dens_rm2}.  In the `no-gas' case, the velocity distribution remains stable, as shown by the overlap of the final velocity distribution (black lines in Fig. \ref{vel_rm2}) with the initial velocity distribution (red lines in Fig. \ref{vel_rm2}).  While the velocity distribution is very mildly dependent upon radius, the inner and outer velocity distributions are still very similar to each other since the density is nearly an isothermal profile.   The density structure is correspondingly stable, remaining in an $\propto r^{-2}$ profile while forming a small $\sim$ 5 M$_{\odot}$ core corresponding to the size of the minimum softening length.

\begin{figure*}
\includegraphics[width=.4\textwidth]{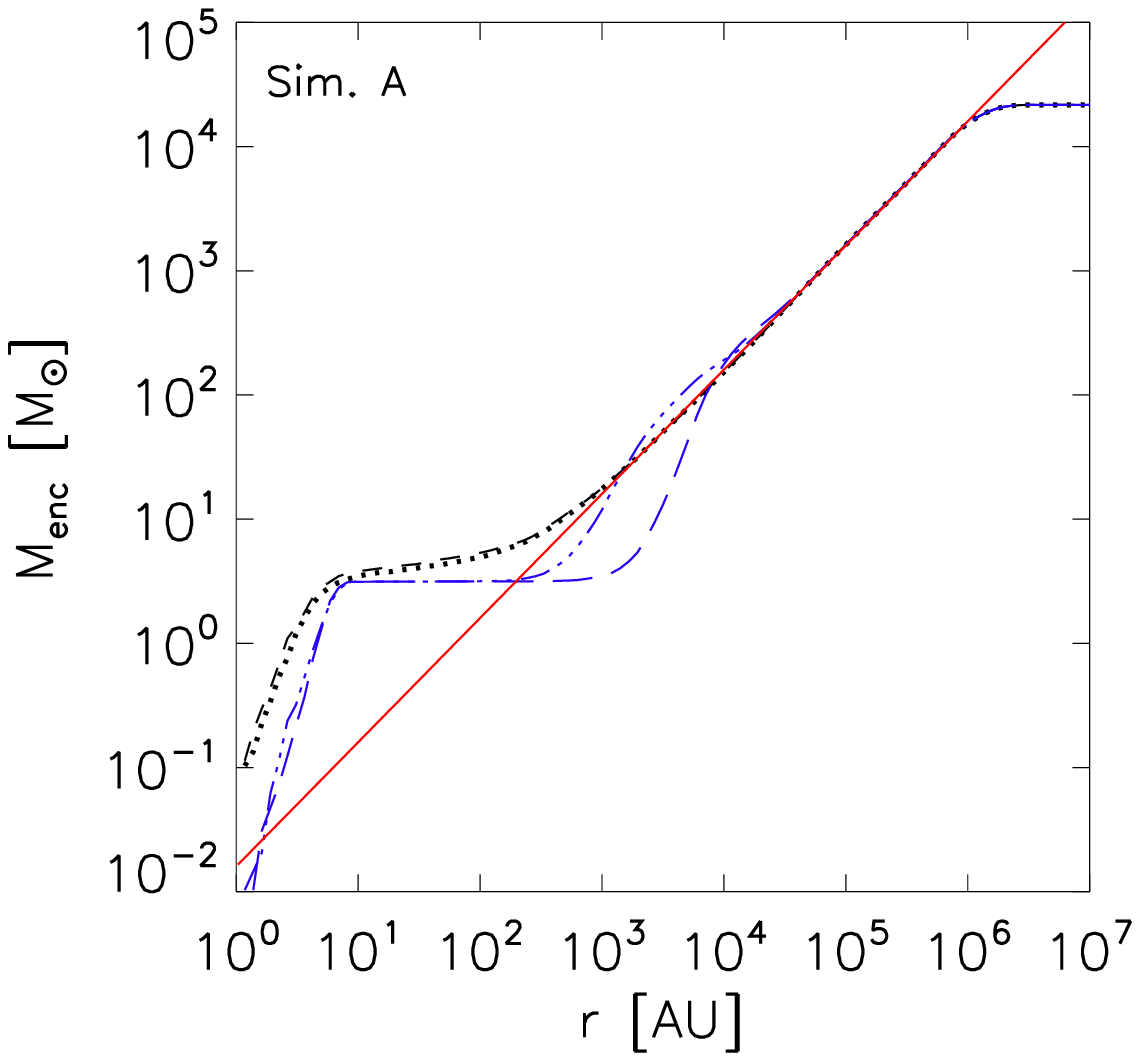}
\includegraphics[width=.4\textwidth]{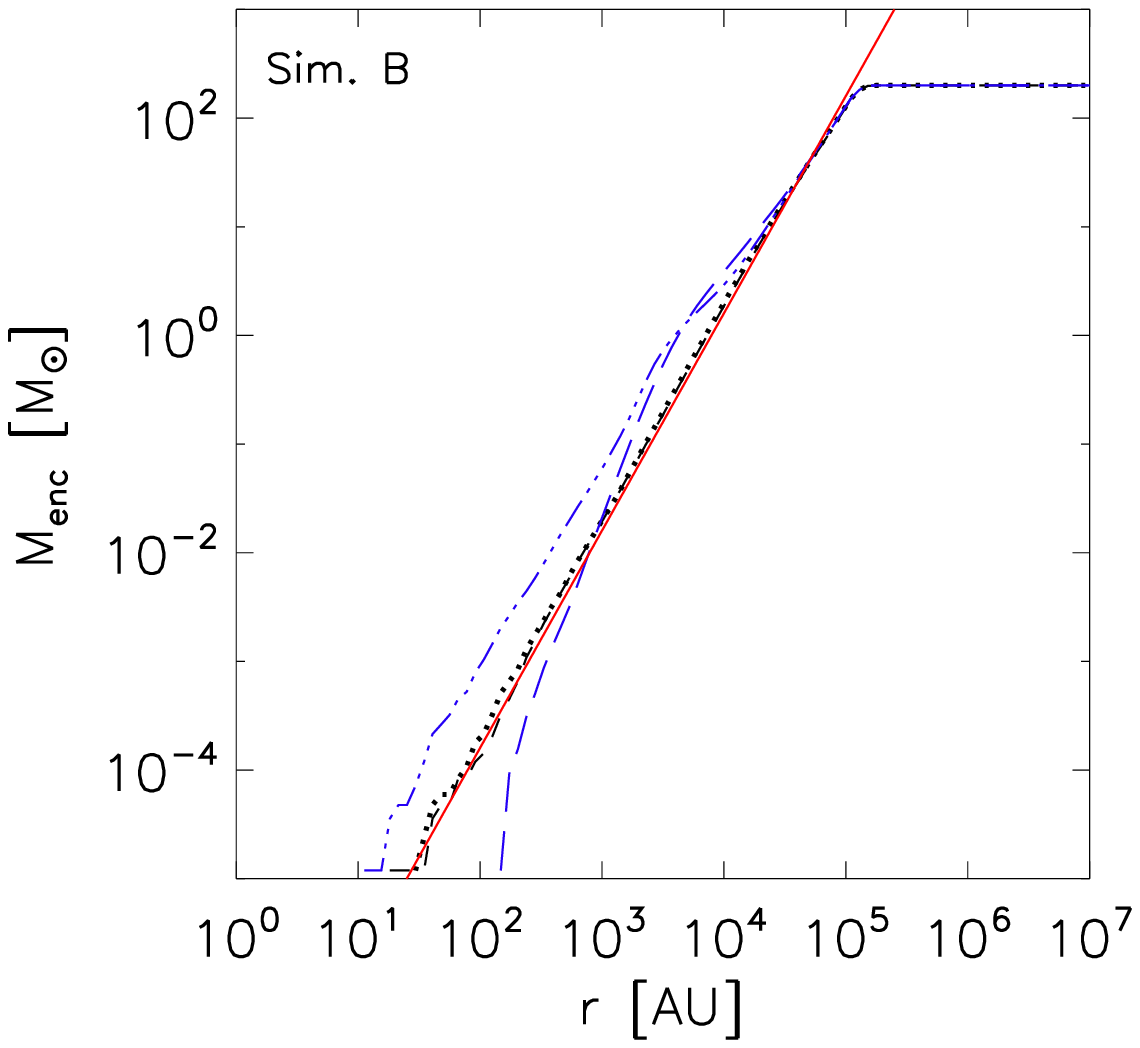}
\caption{ Enclosed DM mass versus radius 
as measured from the densest DM particle.  
Red lines show the analytic fits to which the DM was initialized.  {\it Left:}  `No-gas' case of Simulation A at 50,000 yr (black dotted line), and after a further 20,000 yr (black dashed line).  Note that these lines overlap.  Also shown is the enclosed DM mass for the `with-gas' case after $t_{\rm acc} =$ 10,000 yr (blue dashed triple-dot line) and  $t_{\rm acc} =$ 20,000 yr (blue long-dashed line).  While the enclosed mass within the inner regions slightly increases and stabilizes for the `no-gas' case, the action of  the stellar disk in the `with-gas' case causes the enclosed mass to actually decrease in the inner regions with respect to the evolved `no-gas case' (black), instead becoming relatively more concentrated at farther distances.
 {\it Right}  `No-gas' case of Simulation B at 5 $\times$ 10$^5$ yr (black dotted line), and after a further 20,000 yr (black dashed line).  Note that these lines overlap. Also shown in the enclosed DM mass for the `with-gas' case at $t_{\rm acc} =$ 10,000 yr (blue dashed triple-dot line) and 20,000 yr (blue long-dashed line).  Some gravitational contraction of gas can be seen in the inner 10$^4$ AU at 10,000 yr, but this is insufficient to increase the DM concentration within any of the sinks, and the DM concentration decreases again by 20,000 yr.}
\label{m_enc_rm2}
\end{figure*}

This stability of the density distribution is also shown in Figure \ref{m_enc_rm2}, which displays the mass enclosed within a range of radii at various times in the simulation.  While the `no-gas' case has some initial mass increase in the innermost regions as the core forms, it remains stable thereafter.  This outcome is expected, since the minihalo in the `no-gas' case was initialized to be a in a stable configuration.      
In contrast, the `with-gas' case of Simulation A shows significant evolution due to scattering of the central DM by the star-disk system.  The velocity distribution flattens (Fig. \ref{vel_rm2}), and the central core has slightly less mass than the corresponding region of the `no-gas' case, though the interaction with the gas is insufficient to entirely erase the small DM core.  Fig. \ref{m_enc_rm2} displays how this decrease of enclosed DM mass also occurs on larger scales of a few thousand AU after several thousand yr of interaction with the central gas. 

\begin{figure*} 
\includegraphics[width=.4\textwidth]{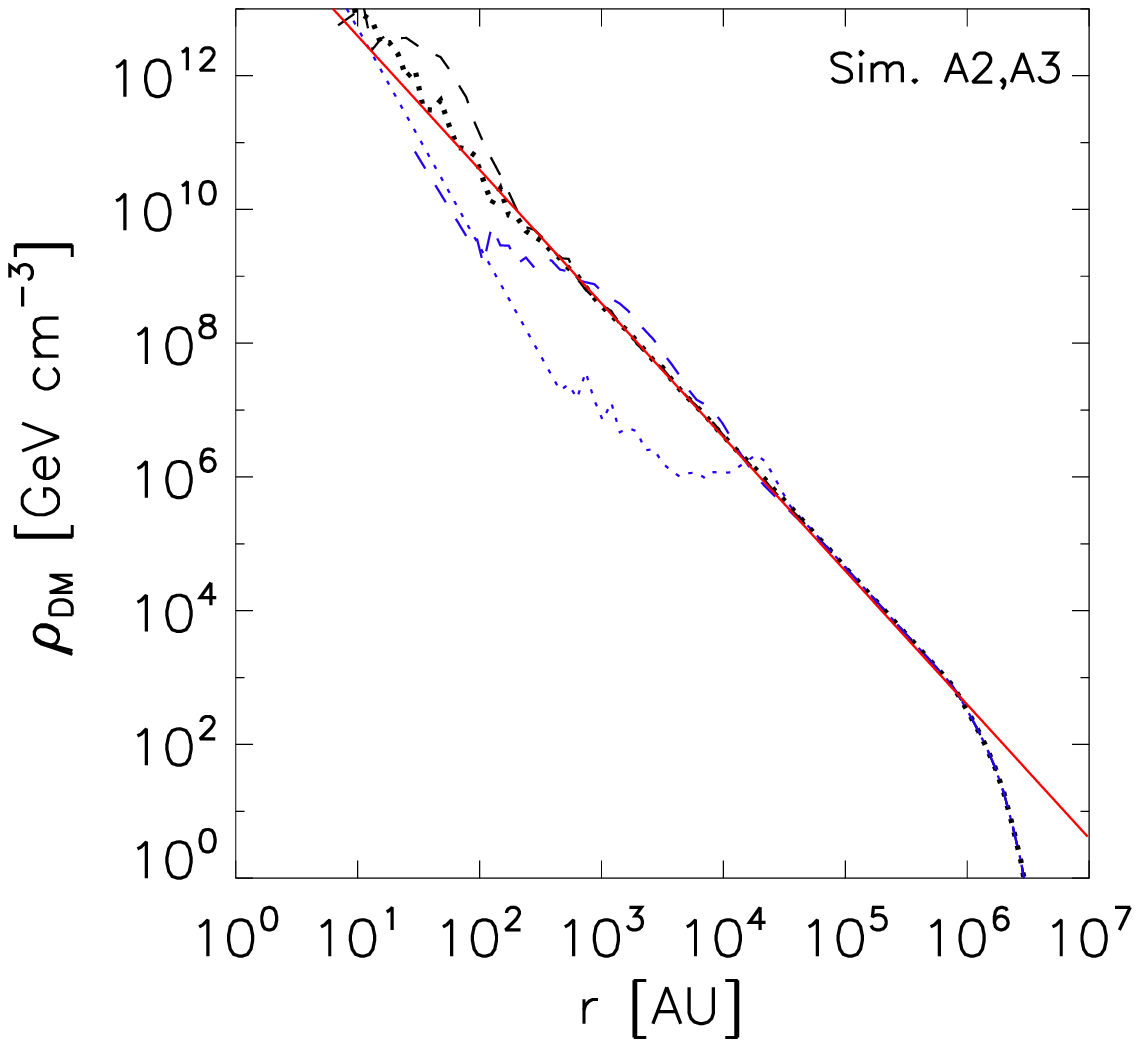}
\includegraphics[width=.4\textwidth]{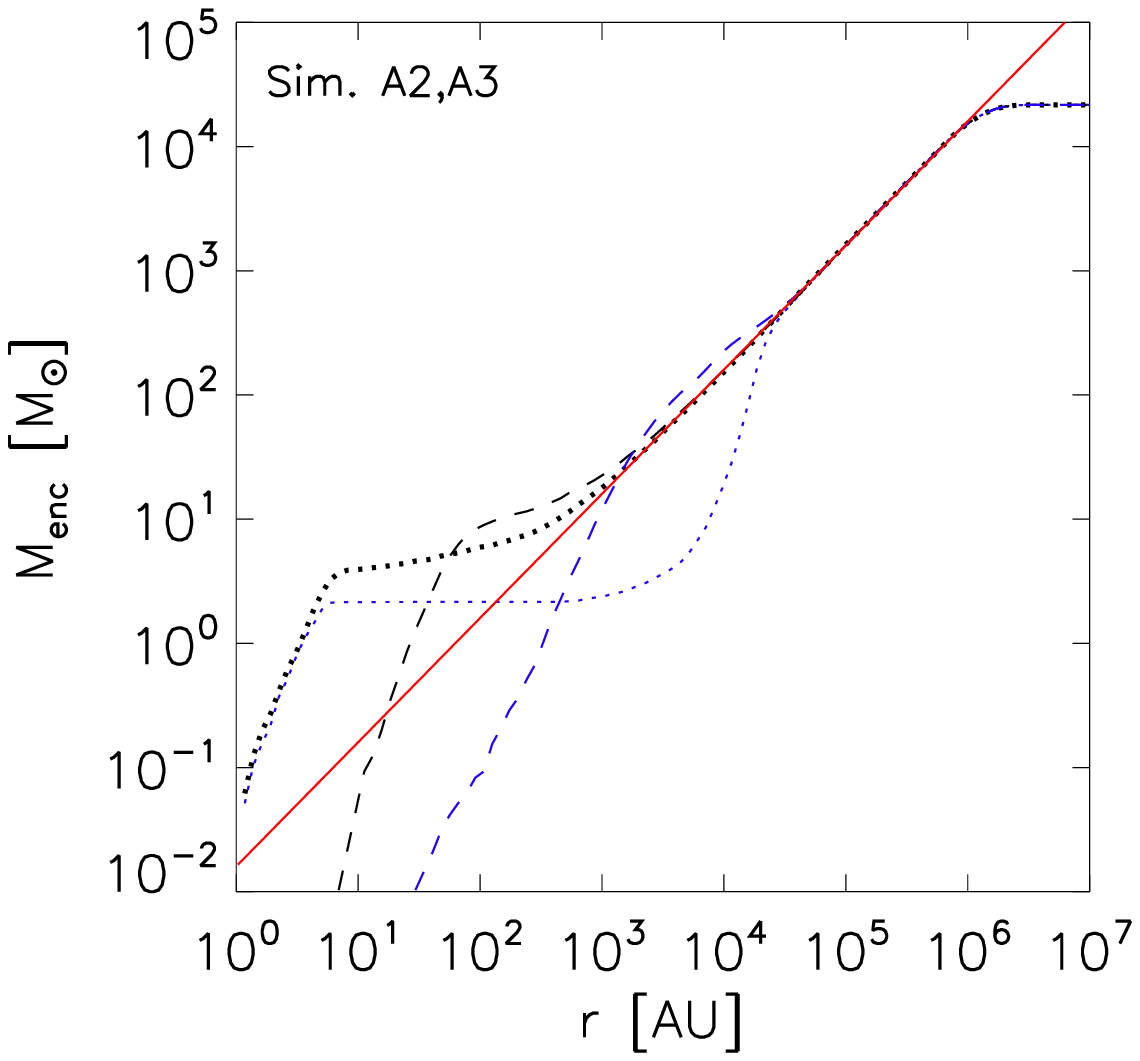}
\caption{Radial distribution of DM at the end of Simulation A2 and A3, which have softening lengths of 5 AU (dotted lines) and 50 AU (dashed lines).  
Black lines are the state of the DM at the end of the `no-gas' cases, after a total of 70,000 yr.  
Blue lines show the DM at the end of the `with-gas' cases, which were each run for 50,000 yr with DM only, and another 20,000 yr after addition of the gas.  
Red lines show the analytical radial profiles to which the DM was initialized.  Left panel shows the density profile, while the right panel shows the profile of enclosed mass.  The initial DM core is largest for the 50 AU case.  Regardless of softening length, however, scattering by the star-disk system shifts most of the central DM to lower densities.}
\label{rm2_test}
\end{figure*}

To determine how the DM evolution in Simulation A varies with resolution, we also performed the same calculation but with a smaller minimum gravitational softening length of 5 AU (Simulation A2), and a larger softening length of 50 AU (Simulation A3).  We find the DM evolution remains largely unchanged, particularly for Simulation A2 (Fig. \ref{rm2_test}).  In the `no-gas' cases a small central core still forms while the density and velocity structures are otherwise stable for over 50,000 yr.  The size of the core scales with the size of the softening length, and thus was largest for Simulation A3.  In each case scattering by the star-disk system lowers the DM density and enclosed mass.  However, this effect is somewhat different for Simulation A3, in which the large softening length allowed the small $\sim$ 10 M$_{\odot}$ core to be completely wiped out by the star-disk system.  In contrast, the corresponding core in Simulations A and A2 remained intact but unaligned with any of the sinks.   

On larger scales between 10$^3$ and 10$^4$ AU, the density and enclosed mass in Simulation A and A2 is actually lower than that in Simulation A3.  Of particular interest is the higher ambient DM density of  10$^9$ GeV cm$^{-3}$ within 10$^3$ AU at the end of Simulation A3.  This is largely because the accretion history of the main sink in Simulation A3 is distinct in that, for the latter half of the  simulation, only a single $\sim$ 30 M$_{\odot}$ sink persisted while the other sinks were lost to mergers.  The lack of multiple sinks weakened the rate at which the surrounding DM  was scattered (see Fig. \ref{rm2_test}), though the disk gas alone managed to significantly lower the surrounding DM density.  
Nevertheless, we find that our results converge for softening lengths of 10 AU or smaller, and even with a larger softening length the qualitative effects of DM on the gas and protostellar evolution will still remain unchanged, as will be discussed in Sections 3.3 and 3.4.

The maximum DM density in Simulation B is much lower, so we initially evolve the DM-only set-up for a longer time of 5$\times$10$^5$ yr ($\sim$ 100 free-fall times for typical peak DM density $\rho_{\rm DM}=10^{8}$ GeV cm$^{-3}$) so that the central DM may still evolve for many free-fall times.  We then add the gas to the resulting DM configuration and again evolve the simulation for $t_{\rm acc} \sim 20,000$ yr (`with-gas' case), and also evolve the DM-only case for another 20,000 yr for comparison (`no-gas' case).  As shown in Figure \ref{vel_rm1}, the DM velocity configuration remains stable for over 5$\times$10$^5$ yr  in the `no-gas' case, while the inner velocity distribution significantly flattens for the `with-gas' case.  
We can also see that, for this DM density profile, the initial velocity distribution differs significantly depending upon radius.
We find that gravitational contraction in fact leads to an increase in DM density within the central 10$^4$ AU over the first $\sim$ 10,000 yr, and the powerlaw of the central profile correspondingly steepens.  However, as shown in Figure \ref{m_enc_rm2} and later in Section 3.3, this gravitational contraction is not sufficient to lead to significant amounts of DM within the sinks themselves.  Scattering by the motion of the star-disk system prevents the central DM from maintaining these enhanced densities, and by 20,000 yr it in fact leads to a decrease again in the central DM densities (Fig. \ref{dens_rm2}).  The enclosed DM mass within the central 1000 AU has decreased as well (Fig. \ref{m_enc_rm2}).

\subsection{Protostellar Mass Growth}

\begin{figure}
 \includegraphics[width=.4\textwidth]{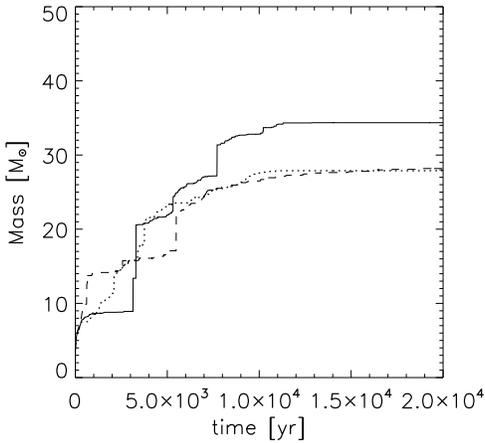}
\caption{Growth of the main sink over time in the various simulations.  Solid line represents the simulation in which the DM was not refined.  Dashed line is taken from Simulation A, and dotted line is the result for Simulation B.  The final masses reached by the largest sink is similar in each of the cases, though a merger at 8000 yr causes a somewhat larger final mass for the `unrefined' case.}
\label{sinkmass}
\end{figure}

We also compare the growth of the sinks without DM refinement (`unrefined') to that found in Simulation A and B (Figure \ref{sinkmass}).
The largest sink was from the `unrefined' simulation, which grew to 34 M$_{\odot}$.  This was still similar to the largest sink masses attained in Simulation A and B, 28 M$_{\odot}$.    
Furthermore, a total of four sinks remained at the end of Simulation A, as compared to three sinks in Simulation B and two sinks in the `unrefined' simulation.  In all three cases the protostellar system is dominated by a massive binary where each member has  mass of $\sim$ 20-30 M$_{\odot}$, and each case furthermore has a total sink mass of $\sim$ 50 M$_{\odot}$.  The presence of refined DM thus had minimal effect on the overall gas accretion rate of the protostars, particularly for the first 5000 yr  ($\sim$ 100 free-fall times for sink density $n=10^{12}$ cm$^{-3}$).  Instead, in all cases the sink accretion rate at early times was similar to that found in, e.g., \cite{stacyetal2010} and \cite{stacyetal2011}.    

\begin{figure*} 
\includegraphics[width=.4\textwidth]{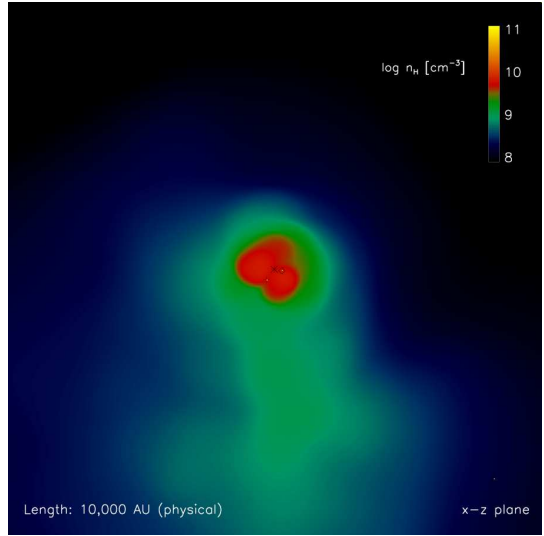}
\caption{Gas density structure of the central 10,000 AU of Simulation A after $t_{\rm acc} =$ 20,000 yr. The main sink is denoted by the asterisk.  Other sinks are denoted by diamonds.  This star-disk system gradually scatters the central DM to progressively lower densities. }
\label{nh_morph}
\end{figure*}

The accretion histories are not all perfectly identical, however.  The early mass growth of the largest sink in each simulation is largely driven by mergers with other sinks, as is evident by the steps in mass growth visible in Figure \ref{sinkmass}.  Thus, even though each simulation began with identical initializations of the central gas, the slight changes in the gravitational potential due to the different DM set-ups were sufficient to cause deviations in the sink merging and growth history.  In particular, a merger at around 8000 yr causes the `unrefined' case to diverge from Simulations A and B.
The sink characteristics are summarized in Table \ref{tab1} for Simulation A, Table \ref{tab2} for Simulation B, and Table \ref{tab3} for the `unrefined' simulation.
An example of the asymmetric, disk-like structure from which the sinks accrete is shown for Simulation A in Figure \ref{nh_morph}.

\begin{table}\begin{tabular}[width=.45\textwidth]{crrrr}
\hline
sink  & $t_{\rm form}$ [yr] & $M_{\rm final}$ [M$_{\odot}$]  & $r_{\rm init}$ [AU] & $r_{\rm final}$ [AU]\\
\hline
1  & 0  & 28  & 0 & 0\\
2  & 740 & 17  & 130 & 260\\
3  & 5540  & 1.0 & 850 & 170\\
4  & 9430  & 0.5 & 660 & 5800\\
\hline
\end{tabular}
\caption{Formation times, final masses, distances from the main sink upon initial formation, and distances from the main sink at the final simulation output in Simulation A. We include the sinks still present at the end of the simulation (20,000 yr).}
\label{tab1}
\end{table}

\begin{table}
\begin{tabular}[width=.45\textwidth]{crrrr}
\hline
sink  & $t_{\rm form}$ [yr] & $M_{\rm final}$ [M$_{\odot}$]  & $r_{\rm init}$ [AU] & $r_{\rm final}$ [AU]\\
\hline
1  & 0  & 28  & 0 & 0\\
2  & 330 & 25  & 66 & 600\\
3  & 6200  & 1.0 & 64 & 1500\\
\hline
\end{tabular}
\caption{Same as Table 1, but for sinks remaining at the end of Simulation B at 20,000 yr.}
\label{tab2}
\end{table}

\begin{table}
\begin{tabular}[width=.45\textwidth]{crrrr}
\hline
sink  & $t_{\rm form}$ [yr] & $M_{\rm final}$ [M$_{\odot}$]  & $r_{\rm init}$ [AU] & $r_{\rm final}$ [AU]\\
\hline
1  & 0  & 34  & 0 & 0\\
2  & 1160 & 16  & 140 &  840\\
\hline
\end{tabular}
\caption{Same as Table 1, but for sinks remaining at the end of the `unrefined' case at 20,000 yr.}
\label{tab3}
\end{table}

\subsection{DM Annihilation Heating on Sink Scales}

\subsubsection{Methodology of Heating Rate Measurement}

We now discuss how we determined the effect of DM annihilation on the central gas. We record the number of DM particles within each sink at every timestep, and  from this we can estimate the DM heating rate.  This rate approximately is:
$\Gamma_{\rm DM} \propto \rho_{\rm DM}^2 \langle \sigma_{a}v\rangle/m_{\rm WIMP}$, where $\langle \sigma_{a}v\rangle$ is the product of the WIMP annihilation cross section and relative velocity, averaged 
over the WIMP momentum distribution, and $m_{\rm WIMP}$ is the mass of a single dark matter WIMP.  $\Gamma_{\rm DM}$ has further dependencies upon the energy spectrum of the photons produced by WIMP annihilation, the scattering of photons within the gas, etc. (see e.g. \citealt{natarajanetal2009}).

We express the evolution of the DM annihilation heating rate $\Gamma_{\rm DM}(t)$
within the sinks in terms of a reference
heating rate $\Gamma_{\rm DM0}$,

\begin{eqnarray}
\frac{\Gamma_{\rm DM}(t)}{\Gamma_{\rm DM0}} 
= \frac{\langle \rho(t)^2\rangle_{\rm sink}}{\langle \rho(t_0)^2\rangle_{\rm sink}} 
&=& \frac{c(t)}{c(t_0)} \frac{\langle \rho(t) \rangle^2_{\rm sink}}{\langle \rho(t_0) \rangle^2_{\rm sink}}
\end{eqnarray}

\noindent where the brackets indicate a spatial average on the scale of the sink
radius. 
We choose the reference time $t_0$ to be the time at which
the average DM density within the main sink reaches its maximum, i.e.,
$\langle \rho(t_0)\rangle_{\rm sink} = \max(\langle \rho(t)\rangle_{\rm sink})$.
The DM clumping factors $c(t) \equiv \langle \rho(t)^2 \rangle_{\rm
sink} / \langle \rho(t) \rangle^2_{\rm sink}$ and 
$c(t_0) \equiv \langle\rho (t_0)^2 \rangle_{\rm sink} / \langle \rho(t_0) \rangle^2_{\rm
sink}$ describe the variance of DM density fluctuations, both
resolved and unresolved, on scales below the sink
particle radius at the respective times.

In our simulations we only track the evolution
of the average DM densities within the sinks, $\langle \rho(t)
\rangle_{\rm
sink}$, at high time resolution. On the other hand, the DM clumping factors
are only known for the few time steps at which a full output of the simulation
data is performed. We therefore estimate the ratio of DM
heating rates using the ratio of the average DM densities 
$\langle \rho(t) \rangle^2_{\rm sink}/ \langle \rho(t_0) \rangle^2_{\rm sink}$. 
This means that we
implicitly ignore that DM clumps
on sub-sink scales.

However, even if DM clumps significantly on sub-sink scales,
computing the ratio of DM heating rates  using only the average DM
densities is a good
approximation if, as one can reasonably expect, the DM clumpiness on
sub-sink scales is
largest when the average DM density within the sink is largest. In
this case, i.e., for $c(t) \le c(t_0)$,
any decrease in the ratio of the
squared average DM densities conservatively underestimates
the decrease in the ratio
of DM heating rates, because $\Gamma_{\rm DM}(t) / \Gamma_{\rm DM0} \le
\langle \rho(t) \rangle^2_{\rm
sink}/ \langle \rho(t_0) \rangle^2_{\rm sink}$. 
Computations of the DM
clumping factors on sub-sink scales
at times where a full data output is available confirm the validity of
this assumption.  

\subsubsection{Evolution of the Heating Rate}

From  \cite{natarajanetal2009} we find that $\Gamma_{\rm DM0}$ is of order 10$^{-8}$ erg cm$^{-3}$s$^{-1}$.  This is the heating rate they found at 20 AU ($10^{-4}$ pc) from the central peak, 
given the typical inner density profiles of their `fit \#2' where $\rho_{\rm DM} \simeq 10^{12}$ GeV cm$^{-3}$ at this distance, and assuming a WIMP mass of $m_{\rm WIMP} = 100$ GeV.  These inner density profiles are 
similar to the initial density profile used in our Simulation A, and we can thus assume a similar heating rate at 20 AU from the DM peak, which initially lies within our main sink.  \cite{natarajanetal2009} find the DM  heating rate will exceed the gas cooling rate within these inner regions, and due to the similar DM construction the same applies to sub-sink gas in Simulation A.  

\begin{figure*}
 \includegraphics[width=.4\textwidth]{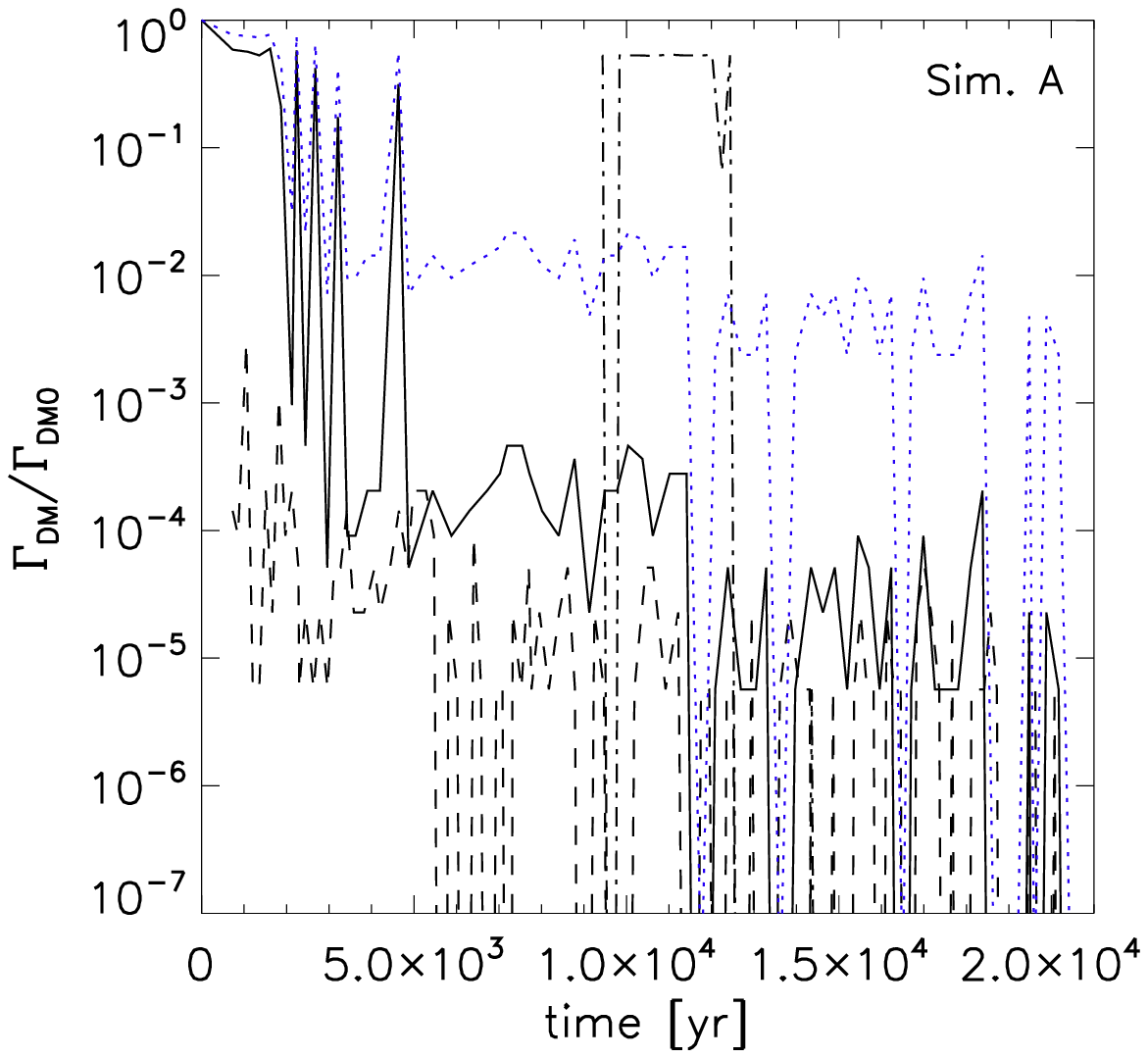}
\includegraphics[width=.4\textwidth]{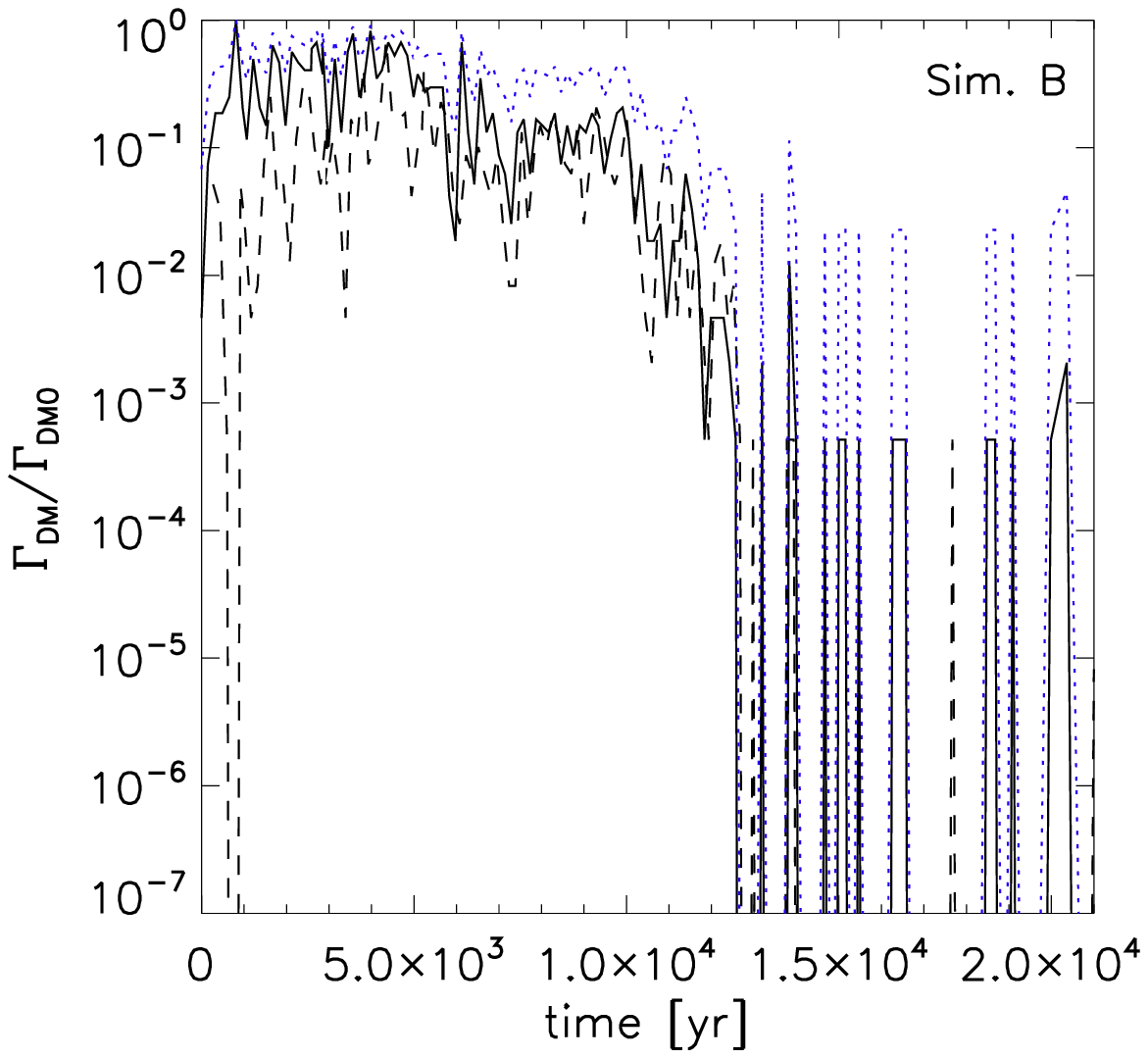}
\caption{The relative DM  heating rate, approximated to be
$\langle \rho(t) \rangle^2_{\rm sink}/ \langle \rho(t_0) \rangle^2_{\rm sink}$ (black lines), 
within the two largest sinks over the 20,000 yr of accretion followed in the simulation.  
The DM capture rate, taken as $\langle \rho(t) \rangle_{\rm sink}/ \langle \rho(t_0) \rangle_{\rm sink}$ (blue dotted lines), is also shown for the largest sink.
Relative heating and capture rates are measured with respect to the highest rate attained within the largest sink.  
{\it Left}: Simulation A ($\Gamma_{\rm DM0} \sim 10^{-8}$ erg cm$^{-3}$s$^{-1}$).  {\it Right:} Simulation B ($\Gamma_{\rm DM0} \ga 10^{-14}$ erg cm$^{-3}$s$^{-1}$).  
Solid lines denote the heating rate for the most massive sink, the dashed lines for second-most massive sink, and dashed-dot line is for a representative small ($\sim $1 M$_{\odot}$) sink.  While there is some oscillation of the DM density and heating rate within the main sink, there is a steadily decline as the star-disk system evolves and disrupts the initial high DM concentration.   $\Gamma_{\rm DM}$  within the second largest sink stays steadily small in Simulation A. 
}
\label{dm_heat}
\end{figure*}

As the DM density profile evolves, however, the heating rate evolves as well. 
As can be seen in Figure \ref{dm_heat}, while the heating rate  $\Gamma_{\rm DM0}$ of \cite{natarajanetal2009} may apply at early times for our Simulation A, $\Gamma_{\rm DM}$ within  $r_{\rm acc}$ of the two largest sinks shows a steady decline over their accretion time as the DM central peak is scattered away.  For the largest sink of Simulation A we see a four order of magnitude drop in heating rate, from 10$^{-8}$ erg\,cm$^{-3}$s$^{-1}$ to 10$^{-12}$ erg cm$^{-3}$s$^{-1}$  by 5000 yr ($\sim$ 1000 free-fall times for typical peak DM density $\rho_{\rm DM}=10^{12}$ GeV cm$^{-3}$),  and to zero by 20,000 yr.   This same zero heating rate was measured at the end of Simulations A2 and A3.  

More precisely, a measured heating rate of zero is simply a lower limit, as the sinks could lie within regions of DM with densities as high as $\sim 10^{10}$ GeV cm$^{-3}$ without the discrete DM particles of the simulation falling within the sink radius, thus yielding a zero measurement when the physical heating is small but non-zero.  Even at $10^{10}$ GeV cm$^{-3}$, however, the annihilation heating rate becomes negligible.  
Furthermore, the distance between the densest DM particle and main sink has grown to $\sim$ 10$^4$ AU by the end of the Simulation A.
Figure \ref{dens_rm2} shows that at these distances the DM densities are much lower than $10^{10}$ GeV cm$^{-3}$ at 20,000 yr,
and were in fact lower than this even at the beginning of the simulation.  Even without any change of the DM profile from its initial configuration, the motion of the sinks from the DM peak was sufficient to severely mitigate DM heating within the sinks.  With the flattening of the DM profile, however, DM heating would not become significant unless the sinks again moved to within $\sim$ 100 AU of the DM peak.  Such realignment is not seen in the simulation, however, due to the orbital motion of star-disk system.

We note that for a smaller WIMP mass, e.g. $m_{\rm WIMP} = 10$ GeV, $\Gamma_{\rm DM}$ will be a factor of ten greater.  In this case the DM heating rate over the first 5000 yr evolves from approximately 10$^{-7}$ erg\,cm$^{-3}$s$^{-1}$ to 10$^{-11}$ erg\,cm$^{-3}$s$^{-1}$ at 20 AU from the sink's center (see discussion in \citealt{natarajanetal2009}).   This roughly corresponds to  
$\Gamma_{\rm DM}$ falling from ten times the gas cooling rate  to 1000 times smaller, given a gas density of 10$^{12}$ cm$^{-3}$.  The cooling rate is even greater for denser gas closer to the center of the sink, while the ambient DM density and heating rate within the sink will continue to decline over time.

\begin{figure*}
 \includegraphics[width=.4\textwidth]{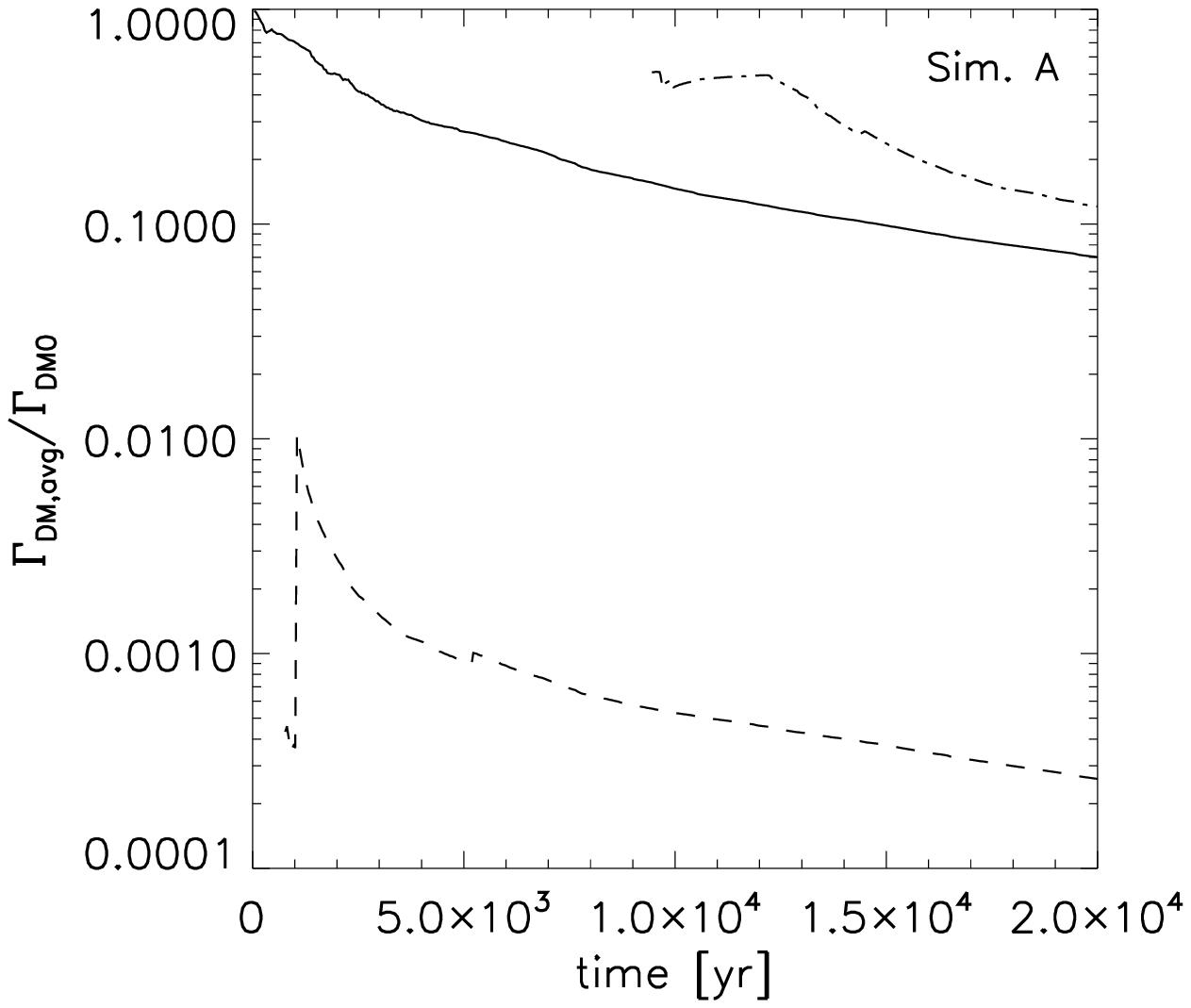}
\includegraphics[width=.4\textwidth]{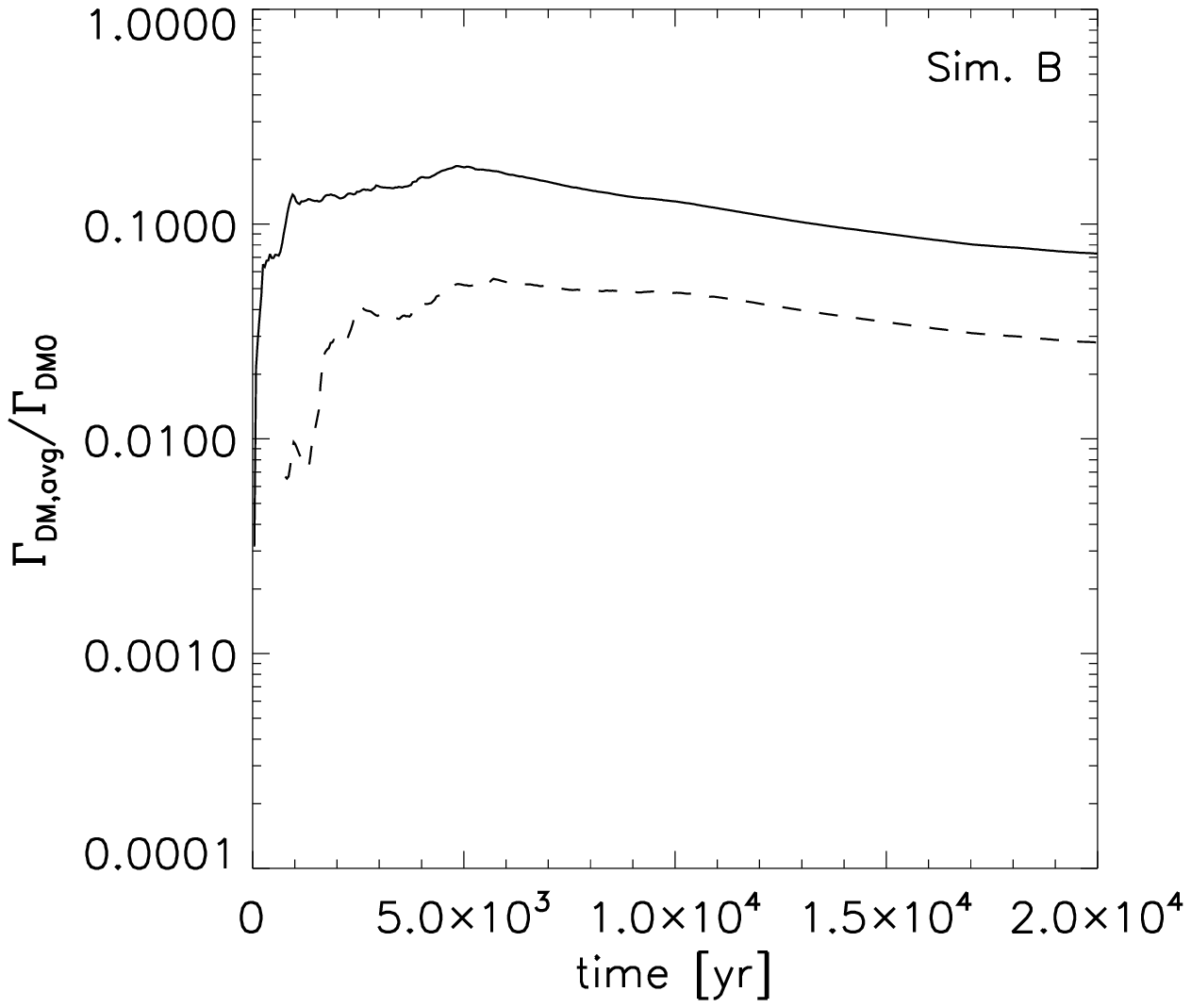}
\caption{Evolution of the time-averaged relative heating rate within the sinks, $\overline{\langle \rho(t_{\rm acc})\rangle^2_{\rm sink}}/ \langle \rho(t_0)\rangle^2_{\rm sink}$, where $\langle \rho(t_0)\rangle^2_{\rm sink}$ is the square of the maximum average density attained by the largest sink of the simulation.   
{\it Left}: is for Simulation A. {\it Right}: Simulation B.  
Solid lines denote the most massive sink, the dashed lines denote the second-most massive sink,  and dashed-dot line is for a representative small ($\sim $1 M$_{\odot}$) sink.  
  The sink and DM density peak are initially aligned.  At early times in Simulation B, gravitational contraction causes a further increase in peak DM density by a factor of $\sim$ 10.  This increase is later reversed through gravitational scattering.
As the amount of sub-sink DM particles declines towards zero in both simulations, $\overline{\langle \rho(t_{\rm acc})\rangle^2_{\rm sink}}$ declines to an order of magnitude below $\langle \rho(t_0)\rangle^2_{\rm sink}$.}
\label{dm_avg}
\end{figure*}

Along with the instantaneous relative heating rate shown in Figure \ref {dm_heat}, we also compare the time-averaged heating rate 
$\Gamma_{\rm DM, avg}$ in Figure \ref{dm_avg}, which for each timestep was calculated as the average of the relative heating rate up to the given time in the simulation, 
$\overline{\langle \rho(t_{\rm acc})\rangle^2_{\rm sink}}/ \langle \rho(t_0)\rangle^2_{\rm sink}$, where 

\begin{equation}
\overline{\langle \rho(t_{\rm acc})\rangle^2_{\rm sink}} =\frac{1}{t_{\rm acc}} \int_0^{t_{\rm acc}} \,  \langle \rho(t)\rangle^2_{\rm sink} \, \mathrm{d} t \mbox{.}
\end{equation}

\noindent Because the presence of DM within the sinks declines until no DM is left in these regions by the end of the simulation, the average heating rate over the first 20,000 yr is an order of magnitude less than its value at the beginning of the simulation. 
A very similar trend of DM heating rates which rapidly decline to a measured zero value is also seen for both smaller and larger resolution scales (Simulations A2 and A3).  

We also note that a few of the sinks formed later in Simulation A initially developed within the persistent $\sim$ 5 M$_{\odot}$ DM core.  This can be attributed to the gravitational enhancement in this region.
A typical example of the evolution of the DM heating rate within such a sink is shown as the dash-dot line in Figure \ref{dm_heat}.  The motion of the sink through the disk causes a separation between the sink and 
DM peak after less than 5000 yr, and the DM heating rate correspondingly declines.  Thus, even when a sink is formed in the location most conducive to DM heating and accretion, large DM heating rates powered by gravitational accretion will be very short-lived compared to the lifetime of the star.  

Even when the softening length was varied in Simulations A2 and A3, the results were quite similar.  The DM heating rate within the main sink dropped three to four orders of magnitude after $\sim$ 10,000 yr, and no DM particles were present within the sink radius by $\sim$ 20,000 yr.  However, the phenomenon of small sinks forming within the small DM core did not occur for the large softening length of Simulation A3, so this may simply be a numerical effect.     
It is furthermore interesting that in Simulation A3, which had only one sink in the latter 10,000 yr, the motion of the disk gas and the single sink alone was sufficient to significantly decrease the influence of DM.  Though there may still have been multiplicity in Simulation A3 on unresolved sub-sink scales, this result shows that disk formation and subsequent gas rotation  can minimize DM effects even without stellar multiplicity.  However, Figures \ref{dens_rm2} and \ref{rm2_test} show that the 
reduction in DM densities between distances of 10$^3$ and 10$^4$ AU from the DM peak is strongest  in the case of a multiple system.

For the largest sink of Simulation A, we thus find that DM heating may be important for $\sim$ 5000 yr, and gravitational accretion of DM will end prior to 20,000 yr.
After this point the DM peak is well over 50 AU away from the protostar, beyond the edge of the sink.  Note that in determining this time estimate we did not account for loss of DM from the 
simulation after it entered the sink region and possibly underwent annihilation.  The total mass of DM within the whole simulation box thus never decreased,
 keeping the central DM potential at a maximum.  We estimate the mass loss of DM through annihilation using the above-quoted heating rate,  10$^{-8}$ erg cm$^{-3}$s$^{-1}$,
which was found assuming $\langle \sigma_{a}v\rangle = 3 \times 10^{-26}$ cm$^3$ s$^{-1}$ (\citealt{natarajanetal2009}).  
 If DM within a sphere of radius 20 AU were lost at this rate over a period of 20,000 yr, this would correspond to a total DM mass loss of 
$\ga 10^{-7}$ M$_{\odot}$.  This is a negligible effect, even considering that the true mass loss may be several times higher than this since annihilation does not heat the gas with perfect efficiency.
The WIMP annihilation cross section is thus unimportant compared to gravitational scattering in determining the evolution of the DM density and heating rate.

For the lower DM densities of Simulation B, typical heating rates within the main sink should be much lower, and we find the heating rates should never become significant to the thermal evolution of the gas.  The initial DM density profile of Simulation B most resembles the shallowest profile of `fit \#1' in \cite{natarajanetal2009}, where $\rho_{\rm DM} \simeq 10^{8}$ GeV cm$^{-3}$ at 20 AU from the peak.  
  In this case $\Gamma_{\rm DM}(t)/\Gamma_{\rm DM0}$ does not start at a value of one because the maximum heating rate $\Gamma_{\rm DM0}$ is not reached until $t_{\rm acc}  \sim 1000$ yr.
  The sink and DM density peak are initially aligned, and gravitational contraction causes the DM peak to increase in density by another factor of $\sim$ 10 for the first 1000 yr.  Given the highest attained density of $\sim 10^{9}$ GeV cm$^{-3}$ at 20 AU from the peak, this corresponds to $\Gamma_{\rm DM0} \ga 10^{-14}$ erg cm$^{-3}$s$^{-1}$ (\citealt{natarajanetal2009}).  Even when the similar DM profile is extrapolated to nearly protostellar scales in \cite{natarajanetal2009}, the heating is still not great enough to influence the gas collapse.  
 Increasing $\Gamma_{\rm DM0}$ a factor of ten by assuming a smaller WIMP mass of $m_{\rm WIMP} = 10$ GeV still yields a heating rate several orders of magnitude lower than the cooling rate of gas within the sink.
  
  As the gas and DM evolve in Simulation B, this remains the case. 
The right panel of Figure \ref{dm_heat} shows that the heating rate within the two largest sinks does not grow after $t_{\rm acc} \sim 1000$ yr.  There is no further increase through, e.g., gravitational contraction.  The growth of $ \langle \rho(t) \rangle_{\rm sink}$ instead halts, and is later reversed.  By 5000 yr $\Gamma_{\rm DM,avg}/\Gamma_{\rm DM0}$ begins a steadily decline (Fig. \ref{dm_avg}).
Similar to Simulation A, the distance between the densest DM particle and main sink has grown to $\sim$ 10$^4$ AU by the end of the simulation.
Thus, for this particular DM configuration, we find DM annihilation heating will not play a significant early role in the formation of massive Pop III stars.  We next discuss the possible longer-term effects of DM scattering accretion.

\subsection{DM Scattering Accretion on Stellar Scales}

 The rate at which DM will be captured by the star through scattering is roughly proportional to the DM density in the vicinity of the star (e.g. \citealt{iocco2008,ioccoetal2008}).  Thus, the DM scattering rate and capture rate will evolve with $\rho_{\rm DM}$, so we roughly calculate the evolution of the capture rate $C$ as
 $C \propto \langle \rho(t) \rangle_{\rm sink}/ \langle \rho(t_0) \rangle_{\rm sink}$ 
 (blue lines in Fig. \ref{dm_heat}).    After 5000-10,000 yr the capture rate drops to 10$^{-2}$ - 10$^{-3}$ of the maximum value, and for both cases the measured value for $C$ drops to zero by 20,000 yr.  We again note this is a lower limit and that $C$ may simply be small but non-zero.  We give further estimates for the true physical values of $C$ below.

While the capture rate will be greatly reduced as the stars move to lower-density regions of DM, we also note that this rate depends on the velocity $v_*$ of the stars themselves with respect to the DM halo.  This dependence on velocity has been calculated by \cite{gould1987} and also discussed in \cite{ioccoetal2008},  though we note a separate study of scattering accretion by \cite{sivertsson&gondolo2011} which accounts for the fact that the concentrated DM is gravitationally bound to the star and assumes that the star does not move through the DM halo.  The equation as determined by \cite{gould1987} is:

\begin{equation}
C = 4 \pi \int_0^{R_*} dR \, R^2 \, \frac{dC(R)}{dV},
\label{Intcaprate}
\end{equation}
\noindent
where

\begin{eqnarray}
\frac{dC(R)}{dV} = \left(\frac{6}{\pi}\right)^{1/2}
\sigma_0 \bar{v} A_n^4 \frac{\rho_\ast}{M_n}\frac{\rho_{\rm DM}}{m_{\rm WIMP}} \mbox{,}
\frac{v^2(R)}{\bar{v}^2} F 
\label{Caprate}\\ \nonumber
\end{eqnarray}

\noindent and $F$ is defined as

\begin{eqnarray}
F & = &
\frac{1}{2\eta A^2}\left\{\left(A_+A_- -\frac12\right)
\left[\chi(-\eta,\eta) -\chi(A_-,A_+)\right]
\right.\nonumber\\
&+&\left.
\frac12A_+e^{-A_-^2} -\frac12A_- e^{-A_+^2}-\eta e^{-\eta^2}
\right\}, 
\end{eqnarray}

\noindent and

\begin{eqnarray}
A^2&=& \frac{3v^2(R)\mu}{2\bar{v}^2\mu_-^2},   \  \  A_\pm=A\pm\eta,  \  \   \eta=\sqrt \frac{3v_\ast^2}{2\bar{v}^2}, \\
\chi(a,b)&=&\int_a^b dy\, e^{-y^2}=
\frac{\sqrt\pi}{2}[{\rm erf}(b)-{\rm erf}(a)] \mbox{.}
\nonumber
\end{eqnarray}

\noindent $R_*$ is the stellar radius, $\sigma_0$ is the WIMP-proton elastic scattering cross section, $A_n$ the atomic number of stellar nuclei, $M_n$ the atomic mass of stellar nuclei, $\bar{v}$ the WIMP velocity dispersion,  $v(R)$ the escape velocity at a given radius $R$ inside the star, $\mu = m_{\rm WIMP}/M_n$, and $\mu_- = (\mu -1)/2$.  As discussed in, e.g., \nocite{freeseetal2008b} Freese et al. 2008b, for $v_* =10 $ km s$^{-1}$, $m_{\rm WIMP} = 100$ GeV, and $\bar{v} =$ 10 km s$^{-1}$, the stellar velocity-dependent part of the capture rate (the factor $F$) amounts to a value of order one.  For example, given a 30 M$_{\odot}$ star that has contracted to 5 R$_\odot$, if the star's motion through the disk leads to a change in $v_*$ from 1 km s$^{-1}$ to 10 km s$^{-1}$,  then F increases from 0.24 to 1.6  This corresponds to an increase in  the capture rate by a factor of $\sim$ 7 due to increased velocity of the star through the DM halo.  \cite{ioccoetal2008} provide a simple estimate for the above capture rate given flat central DM and stellar density profiles, and also assuming that $v_* = \bar{v}  =10 $ km s$^{-1}$, values which apply well to Simulation A.  This leads to their simplified expression:      

\begin{eqnarray}
C &=& 9.2 \times 10^{47} s^{-1} 
\left(\frac{M_*^2}{R_*}\right)
\left(\frac{\rho_{\rm DM}}{10^{11} \rm GeV \, cm^{-3}}\right) \nonumber\\
&\times& \left(\frac{\sigma_0}{10^{-38}\rm cm^{2}}\right)
\left(\frac{m_{\rm WIMP}}{100 \rm GeV}\right)^{-1} \mbox{.}
\end{eqnarray}

\noindent  An initial estimate for the scattering accretion rate can be made with the above equation, in which  $M_*$ is the protostellar mass in solar masses, and $R_*$ is the protostellar radius in cm.  We estimate $M_*=30$ M$_{\odot}$, $R_*=7\times10^{12}$ cm (100 R$_\odot$), $\rho_{\rm DM}=10^{14}$ GeV cm$^{-3}$,  and $m_{\rm WIMP}=100$ GeV.  For $\sigma_0$ we use the spin-dependent cross section of $10^{-38}\rm cm^{2}$, consistent with experimental upper limits for 100 GeV WIMPS as found by, e.g., \cite{desaietal2004} and \cite{angleetal2008}, though we note that more recent results from \cite{behnkeetal2011} find slightly  higher upper limits of $\sim 7 \times 10^{-38}$ cm$^{2}$.  From this we find a capture rate of $10^{41}$ s$^{-1}$.  For 100 GeV DM particles, this corresponds to a total captured DM mass of 10$^{-3}$ M$_{\odot}$ over 5000 yr, and a luminosity from DM annihilation of $L_{\rm DM} \sim 2\times10^{40}$ erg s$^{-1}$  ($= 5\times10^6$ L$_{\odot}$).  A large but brief luminosity, similar to that described in \cite{sivertsson&gondolo2011} despite their different method of calculation, may then be possible.  We again point out, however, that the initial DM capture rate  drops to $\la 10^{-2}$ of the initial value by $\sim$ 10,000 yr.
   This estimate also shows that, because DM is expected to account for only a small fraction of the protostellar mass itself, this effect was safely neglected when following the mass growth of the sinks. 

Over longer periods of time, the stars may continue accreting DM at very reduced rates throughout their main sequence.  This includes the small, $\sim$ 1 M$_{\odot}$, stars that would need less DM to be supported by DM annihilation. There is a $\sim$ 10$^4$ AU DM core of 10$^7$ GeV cm$^{-3}$ at the end of Simulation A,
and a similar core at the end of Simulation B.  
If this core is stable, this would yield a long-term scattering accretion rate of 10$^{33}$~s$^{-1}$ for main-sequence stars with $M_* = 1$~M$_{\odot}$ and $R_* = 1$~R$_\odot$, or a luminosity from DM annihilation of $L_{\rm DM} \sim 2\times10^{32}$ erg s$^{-1}$ ($= 5\times10^{-2}$ L$_{\odot}$), insufficient to power the star.  Larger stars with $M_* =$ 30 M$_{\odot}$ and $R_* =$ 5 R$_\odot$ will have a long-term scattering accretion rate of $2 \times 10^{35}$ s$^{-1}$, and a luminosity of $L_{\rm DM} \sim 4\times10^{34}$ erg s$^{-1}$ (= 10 L$_{\odot}$), again insufficient to power a massive star.   For the higher ambient DM density of  10$^9$ GeV cm$^{-3}$ at the end of Simulation A3, which had a larger 50 AU softening length, this would lead to a luminosity of 10$^3$ L$_{\odot}$ for the single 30~M$_{\odot}$ star in this simulation, still a negligible fraction of the typical main sequence luminosity of such a star  ($\sim 5 \times 10^5$ L$_{\odot}$, e.g. \citealt{hosokawaetal2010}).

Though the precise protostellar evolution is unresolved, and though the contraction of the protostar may indeed be temporarily delayed through DM heating,
Simulation A shows that the initial influence of DM on the protostar will end 
once the star has grown to $\sim$ 30 M$_{\odot}$.  Any DM that may have been incorporated 
onto the protostar early on will be quickly used up as an energy source for the protostar, while later scattering accretion will be minimal.   \cite{natarajanetal2009} estimate the initial DM supply will be exhausted by $\sim 10^5$ yr, 
after which the protostar will subsequently undergo Kelvin-Helmholtz contraction onto the main sequence as usual.  More precisely, they estimate the depletion of  $t_{\rm dep}$ of DM to be

\begin{eqnarray}
 t_{\rm dep}&=& \frac{\rho_{\rm DM}}{\dot{\rho}_{\rm DM}}
 \simeq \frac{m_{\rm WIMP}}{\rho_{\rm DM} \langle \sigma_{a}v\rangle}  \nonumber\\
 &=& 10^2 {\rm Myr} \left( \frac{m_{\rm WIMP}}{100 \rm GeV}  \right)
 \left(\frac{\rho_{\rm DM}}{10^{12} \rm GeV \, cm^{-3}}\right)^{-1} 
 \mbox{.}
\end{eqnarray}
 
 \noindent Compressing the total accreted DM mass calculated above, 10$^{-3}$ M$_{\odot}$, to the size of the protostar (100 R$_\odot$) would yield a slightly higher DM density of $\rho_{\rm DM}=10^{15}$ GeV cm$^{-3}$.  This would indeed correspond to a depletion timescale of $\sim 10^5$ yr, significantly shorter than the expected Pop III lifetime of several Myr.  If we determine a powerlaw fit to the latter 10,000 yr of mass growth of Simulation A's largest sink, we find $M_* \propto t^{.071}$.  Extrapolating to $\sim 10^5$ yr, we find that the star will have only grown to 32 M$_{\odot}$ by the time the DM is depleted and the protostar can contract to the main sequence.  Even if this depletion time were increased by a factor of ten such that  $t_{\rm} \simeq 10^6$ yr, the star will only grow to $\sim$ 40 M$_{\odot}$ in this time, given the same extrapolation.  With so little late-time DM scattering accretion, DM annihilation of the small amount of mass gathered early on is therefore unlikely to extend the gas accretion time of the typical Pop III star long enough to significantly alter the final stellar mass.  

While the stars should spend most of their time in the large low-density DM core, we have thus far neglected the possibility that the sinks will re-align with the smaller, high-density DM core (see Fig. \ref{dens_rm2}) at some points during their orbit through the disk.  Were a $\sim$ 100 AU region of 10$^{10}$ GeV cm$^{-3}$ to persist, a star going at an average relative speed of 10 km s$^{-1}$ would pass through the dense region in 50 yr, very briefly enhancing the luminosity up to  50~L$_{\odot}$ and 10$^4$~L$_{\odot}$ for a 1 M$_{\odot}$ and 30 M$_{\odot}$ star, respectively.  This is less than one-tenth of the main-sequence luminosity of the most massive stars in our simulations, but is much more significant for stars of $\sim$ 1 M$_{\odot}$.  However, considering the stellar orbits through the disk typically cover a radius of 1000 AU, the area covered by $> 10^{10}$ GeV cm$^{-3}$ DM relative to the stellar orbital area is a factor of 10$^{-2}$, so periods of high DM luminosity should not represent a significant portion of the stellar lifetime.

\section{Discussion and Conclusion}

We performed multiple N-body and SPH simulations with {\sc gadget2} to examine the mutual interaction between Pop III stellar disks and central DM cores. 
Our results are highlighted in Figure \ref{dm_heat}.
We find that the motion of the Pop III star-disk system will cause a flattening of the central DM peak.  This also causes a separation between the location of the sinks and the DM peak
of $\sim$ 10$^4$ AU
, so that after a few thousand years DM will not be dense enough to have an effect on the evolution the gas and protostars within the sinks 
(Figs. \ref{dens_rm2} and \ref{dm_heat}).   
 For the steeper profile of Simulation A, we find that even if the DM density profile were unaltered from its initial set-up, the relative distance between the sinks and the DM peak is large enough that the DM will be too diffuse to affect the sub-sink gas and stellar evolution.  The observed flattening of the profile, however, allows the sinks to be as close as $\la$ 100 AU from the DM peak while still avoiding significant DM influence, though such close distances are not seen in the latter part of the simulation.  DM is even less influential in the shallower profile of Simulation B.
 Our results show that Pop III stars will  grow and contract to the main sequence largely unaffected by DM, and the final stellar masses will most likely reach only a few tens of solar masses.

Previous studies of Pop III dark stars have not considered their formation in the context of a disk and stellar multiple system, and none of these studies have utilized three-dimensional calculations.  Other one-dimensional studies by, e.g., \cite{freeseetal2008} and \cite{spolyaretal2009} found that, given a stationary star in the center of a minihalo, gravitational accretion of DM will be prolonged, providing a sufficient power source to delay protostellar collapse for $\sim$ 10$^6$ yr.  This is in contrast to the abbreviated period of DM influence found in our scenario of a stellar multiple system ($\la$ 10$^4$ yr).   \cite{ioccoetal2008}, on the other hand, find that collapse will be delayed by gravitational accretion of DM only for 
$\sim$ 10$^3$ - 10$^4$ yr, more consistent with the timescales found in our work,
though their calculation did not account for continued gas infall onto the star.

Perhaps even more crucial, however, is the rate and prolongation of scattering accretion of DM after the protostar has begun contraction to the main sequence.  Previous studies (e.g. \nocite{freeseetal2008b} Freese et al. 2008b, \citealt{yoonetal2008,ioccoetal2008, spolyaretal2009}) have suggested that dark stars may survive indefinitely through scattering accretion if the surrounding DM medium remains sufficiently dense, usually $\rho_{\rm DM} > 10^{10} - 10^{11}$ GeV cm$^{-3}$.  These studies were unable to address how long such high densities could survive, however, as they could not address the evolution of the DM structure under halo mergers, further star formation, etc.  We find, however, that within a Pop III multiple star-disk system the ambient DM density will rapidly decline, such that $\rho_{\rm DM}$ becomes too low to sustain Pop III stars after only a few thousand yr (Fig. \ref{dens_rm2} ).   

We note that we did not include a DM heating term in our simulations, and we initialized the simulations only after the gas had already collapsed to high density.  Thus, we did not address how DM heating affects the initial baryon collapse before reaching sink densities.  However, this has also been addressed with one-dimensional calculations (e.g. \citealt{spolyaretal2008, natarajanetal2009,ripamontietal2010}).  These studies all found that DM annihilation heating will eventually dominate cooling at  high densities ($n \sim 10^{12}$ cm$^{-3}$), scales unresolved in our study.  Earlier work assumed that this would halt the protostellar contraction and extend the protostellar accretion time.   However, the most recent and detailed study by \cite{ripamontietal2010} finds that the collapse and thermal history of the gas will be relatively unaffected even under the dominance of DM heating.   
Nevertheless, performing similar calculations in three-dimensions is an important task for future work.  It will furthermore be important to self-consistently follow the contraction of DM to small, nearly protostellar scales, particularly to better constrain the inner DM density profile.  We could only parameterize this in the current calculation.  The results we present here, however, demonstrate that even if \cite{ripamontietal2010} have underestimated the typical effect of DM on the collapse of gas to high densities, the multiplicity of Pop III stars leads to yet another way that the influence of DM will be rapidly mitigated. 

Our results are also consistent with constraints to the reionization history provided by WMAP and observations of the Gunn-Peterson trough, since a population of highly massive dark stars  ($\sim$ 1000 M$_{\odot}$) only fits these constraints under a particular scenario of double-reionization (\citealt{schleicheretal2009}).  
On the other hand, \cite{scottetal2011} find that highly massive dark stars can be made consistent with our knowledge of reionization history by modifying other parameters of reionization models such as the star formation efficiency within haloes and the escape fraction of ionizing radiation. 
As discussed in, e.g., \cite{schleicheretal2009} and \cite{yuanetal2011}, a lack of such a dark star population also relaxes constraints on models and rates of DM annihilation set by the x-ray,  $\gamma$-ray, and neutrino observations in the cosmic background and in the Milky Way (e.g. \citealt{ullioetal2002,knodlsederetal2003,beacometal2007,yukseletal2007,macketal2008}).

In addition, our results easily fall in line with further constraints discussed in \cite{zackrissonetal2010b}, who find that current survey data, including that from the {\it Hubble Space Telescope}, already reveals that $10^7$ M$_{\odot}$ dark stars at $z \sim 10$ must be exceedingly rare and also unlikely to be observed by the {\it James Webb Space Telescope}.  However, lower-mass dark stars are still not ruled out by observations, and {\it JWST} may be able to detect these through the aid of gravitational lensing by foreground galaxy clusters (\citealt{zackrissonetal2010}).

The disruption of central DM cusps on larger scales by baryonic motion and stellar feedback  has been studied extensively (e.g. \citealt{elzantetal2001,gnedin&zhao2002,mashchenkoetal2006,governatoetal2010,coleetal2011}).  Earlier work has found that the transfer of energy from the baryons to DM through gravitational interaction can indeed flatten the central DM density profile, and our work shows similar results on stellar disk scales.  In short, we expect that previous studies of Pop III star formation lost little accuracy due to exclusion of DM annihilation effects at high densities. 
We furthermore do not expect to detect either low-redshift Pop III stars whose lifetimes have been extended through accretion of DM or Pop III stars that have grown to extremely large masses due to a gas accretion phase that was extended through DM annihilation effects.
This is because the influence of DM on Pop III  stars will already end well before the stars have reached more than a few tens of solar masses.  

\section*{Acknowledgments}

The authors thank Aravind Natarajan, Fabio Iocco, and Tanja Rindler-Daller for helpful discussions.
A.S. is grateful for support from the NASA Postdoctoral Program
(NPP).  
V.B. acknowledges support from NSF grant AST-1009928, NASA ATFP grant NNX09AJ33G, and JPL Research Support
Agreement 1354840.  This work was supported in part by NSF grant AST-0907890 and NASA grants
NNX08AL43G and NNA09DB30A (for A.L.).   V.B. thanks the Max-Planck-Institut f\"{u}r Astrophysik for its hospitality during part of the work on this paper.
The simulations were carried out at the Texas Advanced Computing Center (TACC).

\section*{Appendix A: Drawing From the Distribution Function}

We here describe in more detail how we randomly draw a value of relative energy $\mathcal E$ from the distribution function  $f(\mathcal E)$.  For a given density profile $\rho(r)$ we generate a function $\Psi(r)$, where $r$ is the distance from the center of the spherically symmetric DM structure.  We set

\begin{equation}
\Psi(r) = -\Phi(r) + \Phi_0  \mbox{,}
\end{equation}

\noindent where $\Phi$ is the gravitational potential and $ \Phi_0$ is a normalizing constant.  We determine $\Phi(r)$ over the range $0 < r < R$, where $R$ is the radius of the entire DM structure, and

\begin{equation}
\Phi(r) = - 4 \pi G \, \frac{1}{r}\int_0^r \rho(r') r'^2 \, \mathrm{d} r'  \mbox{.}
\end{equation}

\noindent We also set

\begin{equation}
\Phi_0 = - 4 \pi G \, \frac{1}{R}\int_0^R \rho(r') r'^2 \, \mathrm{d} r'  
\end{equation}

\noindent so that $\Psi(r=R)$ is zero.  Once $\Psi$ is known for the relevant range of radii and densities, this can be plugged into the expression for $f(\mathcal E)$.  We compute $f(\mathcal E)$ over 3000 logarithmically spaced values of  $\mathcal E$, from $0 \le \mathcal E \le \Psi_{\rm max}$, where $ \Psi_{\rm max} = -\Phi_0$.  

When assigning velocities to each particle, we first determine the value for $\Psi$ based upon the particle's radial distance $r$ from the center of the DM structure.  For the range of $\mathcal E$ values we furthermore determine the probability $P(\mathcal E \le \mathcal E_0)$ that the DM particle will have a relative energy $\mathcal E \le \mathcal E_0$ for the given $\Psi$.  

Recall from \cite{binney&tremaine2008} that we are defining the distribution function such that

\begin{equation}
\rho(r)  = \int f(r,v) \, \mathrm{d}^3 v \mbox{.}
\end{equation}

\noindent Because we have a  spherically symmetric system, and given that $v = \sqrt{2(\Psi - \mathcal E)}$, we can also write 

\begin{eqnarray}
\rho(\Psi)  &=& 4 \pi \int f(\Psi,v) \, v^2 \, \mathrm{d} v \nonumber \\
&=& 4 \pi \int_0^{\Psi} f(\mathcal E') \sqrt{2(\Psi - \mathcal E')} \,\mathrm{d}\mathcal E' \mbox{.}
\end{eqnarray}

From this we derive the following expression for determining $P(\mathcal E \le \mathcal E_0)$:

\begin{equation}
P(\mathcal E \le \mathcal E_0) =   \frac{4 \pi}{\rho(\Psi)} \int_{0}^{\mathcal E_0}  f(\mathcal E')\sqrt{2(\Psi - \mathcal E')} \,  \mathrm{d} {\mathcal E'} \mbox{.}
\end{equation}

\noindent  $P(\mathcal E  \le \mathcal E_0)$ is only considered over the range $0 \le \mathcal E \le \Psi$, such that $P(\mathcal E > \Psi) = 0$.   

We furthermore normalize $P(\mathcal E  \le \mathcal E_0)$ according to $P_{\rm norm}(\mathcal E \le \mathcal E_0) = P(\mathcal E \le \mathcal E_0)/ N(\Psi)$, where

\begin{equation}
N(\Psi) =   \frac{4 \pi}{\rho(\Psi)} \int_0^{\Psi}  f(\mathcal E')\sqrt{2(\Psi - \mathcal E')}  \, \mathrm{d} {\mathcal E'} \mbox{.}
\end{equation}

\noindent The $N({\Psi})$ normalization factor ensures that the total probability over $0 \le \mathcal E \le \Psi$ will sum to one.  

With $P_{\rm norm}$ determined, we next generate a random number  $p_{\mathcal E}$  between 0 and 1.  Finally, we find the value of relative energy $\mathcal E_0$ such that $P_{\rm norm}(\mathcal E \le \mathcal E_0) = p_{\mathcal E}$.  The particle is set to this $\mathcal E_0$, and velocity of the particle is then determined as described in Section 2.2.   

\bibliographystyle{mn2e}
\bibliography{ds}

\label{lastpage}

\end{document}